\documentclass[manuscript=article]{achemso}
\usepackage{xcolor}
\usepackage{float}
\DeclareUnicodeCharacter{2212}{-}

\setkeys{acs}{usetitle=true}

\usepackage{caption}
\usepackage{subcaption}
\usepackage{amsmath}
\usepackage[version=3]{mhchem}

\title[Robust Training of Machine Learning Interatomic Potentials with Dimensionality Reduction and Stratified Sampling]{Robust Training of Machine Learning Interatomic Potentials with Dimensionality Reduction and Stratified Sampling}
\author{Ji Qi}
\affiliation[UCSD]{Materials Science and Engineering Program, University of California San Diego, 9500 Gilman Dr, Mail Code 0448, La Jolla, CA 92093-0448, United States}
\author{Tsz Wai Ko}
\affiliation[UCSD]{Department of NanoEngineering, University of California San Diego, 9500 Gilman Dr, Mail Code 0448, La Jolla, CA 92093-0448, United States}
\author{Brandon C. Wood}
\affiliation[LLNL]{Quantum Simulations Group, Lawrence Livermore National Laboratory, Livermore, CA, 94550, USA}
\alsoaffiliation[LLNL]{Laboratory for Energy Applications for the Future (LEAF), Lawrence Livermore National Laboratory, Livermore, CA, 94550, USA}
\author{Tuan Anh Pham}
\email{pham16@llnl.gov}
\affiliation[LLNL]{Quantum Simulations Group, Lawrence Livermore National Laboratory, Livermore, CA, 94550, USA}
\alsoaffiliation[LLNL]{Laboratory for Energy Applications for the Future (LEAF), Lawrence Livermore National Laboratory, Livermore, CA, 94550, USA}
\author{Shyue Ping Ong}
\email{ongsp@ucsd.edu}
\affiliation[UCSD]{Department of NanoEngineering, University of California San Diego, 9500 Gilman Dr, Mail Code 0448, La Jolla, CA 92093-0448, United States}
\date{}

\begin{document}

\maketitle

\begin{abstract}
Machine learning interatomic potentials (MLIPs) that enable accurate simulations of materials at scales beyond conventional first-principles approaches have played increasingly important roles in understanding and design of materials. However, MLIPs are only as accurate and robust as the data they are trained on. In this work, we present DImensionality-Reduced Encoded Clusters with sTratified (DIRECT) sampling as an approach to select a robust training set of structures from a large and complex configuration space. By applying DIRECT sampling on the Materials Project relaxation trajectories dataset with over one million structures and 89 elements, we develop an improved materials 3-body graph network (M3GNet) universal potential that extrapolate more reliably to unseen structures. We further show that molecular dynamics (MD) simulations with universal potentials such as M3GNet can be used in place of expensive \textit{ab initio} MD to rapidly create a large configuration space for target materials systems. For demonstration, we combined this scheme with DIRECT sampling to develop a reliable moment tensor potential for titanium hydrides without the need for iterative optimization. This work paves the way towards robust high throughput development of MLIPs across any compositional complexity.
\end{abstract}

\section{Introduction}

Machine learning interatomic potentials (MLIPs) have become an indispensable staple in the computational materials toolkit. MLIPs parameterize the potential energy surface (PES) of an atomic system as a function of local environment descriptors using ML techniques.\cite{behlerGeneralizedNeuralNetworkRepresentation2007, bartokGaussianApproximationPotentials2010,thompsonSpectralNeighborAnalysis2015,shapeevMomentTensorPotentials2016, wangDeePMDkitDeepLearning2018, chenUniversalGraphDeep2022,batznerEquivariantGraphNeural2022, NEURIPS2022_4a36c3c5,gasteigerGemNetUniversalDirectional2021,dengCHGNetPretrainedUniversal2023} While MLIPs generally exhibit much better accuracies in energies and forces compared to traditional IPs,\cite{zuoPerformanceCostAssessment2020, unkeMachineLearningForce2021} their key advantage is that they can be systematically fitted and improved in a semiautomated fashion for diverse structural and chemical spaces. By enabling accurate and efficient simulations over length and time scales much larger than those accessible by \textit{ab initio} methods, MLIPs have provided new insights into a wide range of physicochemical processes. These include lithium diffusion in lithium superionic conductors and their interfaces,\cite{qiBridgingGapSimulated2021, leeAtomicscaleOriginLow2023, holekevichandrappaThermodynamicsKineticsCathode2022, winterSimulationsMachineLearning2023, wangFrustrationSuperIonic2023} dislocation behavior and ordering in multiple principal element alloys,\cite{liComplexStrengtheningMechanisms2020, yinAtomisticSimulationsDislocation2021} liquid–amorphous and amorphous–amorphous transitions in silicon,\cite{deringerOriginsStructuralElectronic2021} and reaction mechanisms of molecule-molecule and molecule-surface scattering,\cite{riveroReactiveAtomisticSimulations2019, liuConstructingHighDimensionalNeural2018} to name a few.\cite{unkeMachineLearningForce2021} 

An exciting recent innovation in MLIPs is graph deep learning architectures.\cite{chenUniversalGraphDeep2022,batznerEquivariantGraphNeural2022, NEURIPS2022_4a36c3c5,gasteigerGemNetUniversalDirectional2021,dengCHGNetPretrainedUniversal2023} Graph deep learning models encode the elemental character of each atom using features with a fixed dimensionality, avoiding the combinatorial explosion in model complexity associated local environment descriptors with number of elements. Of particular relevance to this work is the Materials 3-body Graph Network (M3GNet) architecture, which combines many-body features of traditional IPs with those of flexible material graph representations. By training on the massive database of structural relaxations in the Materials Project, \citet{chenUniversalGraphDeep2022} have developed a M3GNet universal potential (M3GNet-UP) for 89 elements of the periodic table and demonstrated its application in predicting structural and dynamical properties for diverse materials. 

The critical challenge in developing a robust MLIP is generating a training dataset that can provide a good coverage of the structural/chemical space of the materials of interest (henceforth, referred to as the ``configuration space''). Typically, the configuration space is generated through domain expertise, comprising ground-state structures, relaxation trajectory snapshots, strained structures, \textit{ab initio} MD (AIMD) structures, defect structures, etc. \textit{Ab initio} calculations such as those based on density functional theory (DFT) are then performed on structures sampled from the configuration space to obtain accurate energies and forces as training data for MLIPs. 

To ensure sufficient coverage of the configuration space, state-of-the-art training protocols often incorporate some form of active learning (AL).\cite{artrithHighdimensionalNeuralNetwork2012, podryabinkinActiveLearningLinearly2017, gubaevAcceleratingHighthroughputSearches2019, zhangActiveLearningUniformly2019, vandermauseOntheflyActiveLearning2020, sivaramanMachinelearnedInteratomicPotentials2020} In this way, an MLIP is used to simulate the materials of interest, and generated structures that require extrapolation are added to refit the MLIP in an iterative fashion. The key advancement in AL is the efficient uncertainty evaluation of MLIP prediction on new structures without referring to the DFT PES, which greatly expands the search space and minimizes the cost of training structure augmentation. While AL has been undeniably effective in the construction of robust MLIPs, it can be inefficient for highly complex configuration spaces. For instance, a recent work by the authors to fit a moment tensor potential for the 7-element \ce{(Li_{7/18}Sr_{17/36})(Ta_{1/3}Nb_{1/3}Zr_{2/9}Sn_{1/9})O3 } complex concentrated perovskite required over 100 AL iterations.\cite{koCompositionallyComplexPerovskite2022} 

An ideal strategy should enable efficient generation and sampling of the configuration space prior to any DFT computations. One proposed approach is to bias MD simulations to sample ordered and disordered structures as an entropy maximization (EM) strategy\cite{karabinEntropymaximizationApproachAutomated2020, montesdeocazapiainTrainingDataSelection2022} to sample a diverse feature space. For example, \citet{montesdeocazapiainTrainingDataSelection2022} showed that an MLIP for tungsten trained with an EM set has much more consistent accuracies in energies for structures present in both EM set and domain expertise (DE) training set, while the MLIP trained with DE set performs significantly worse for EM set than for DE set. Another recently proposed high-throughput scheme generated four training sets for Mg, Si, W and AL by applying normally distributed random atom displacements together with isotropic and anisotropic lattice scaling to the respective non-diagonal supercells.\cite{lloyd-williamsLatticeDynamicsElectronphonon2015, allenOptimalDataGeneration2022} The as-fitted MLIPs can accurately reproduce the force constant matrix of those crystalline systems.

In this work, we present a DImensionality-Reduced Encoded Clusters with sTratified (DIRECT) sampling strategy to generate robust training data for MLIPs for any chemical systems.  We will first demonstrate the effectiveness of DIRECT sampling of 1.3 million structures in the Materials Project structural relaxation dataset\cite{chenUniversalGraphDeep2022, chenchiMPF20212022, jainCommentaryMaterialsProject2013} to fit an improved M3GNet universal potential (UP). Next, we will demonstrate how the M3GNet UP can be used to effectively generate configuration spaces for DIRECT sampling using the Ti-H model system, which is known to be highly challenging for reliable MD simulations. This work paves the way towards robust high throughput development of MLIPs across any
compositional complexity.

\section{Results}

\subsection{DIRECT Workflow}
\begin{figure}[H]
    \centering
    \includegraphics[width=0.8\textwidth]{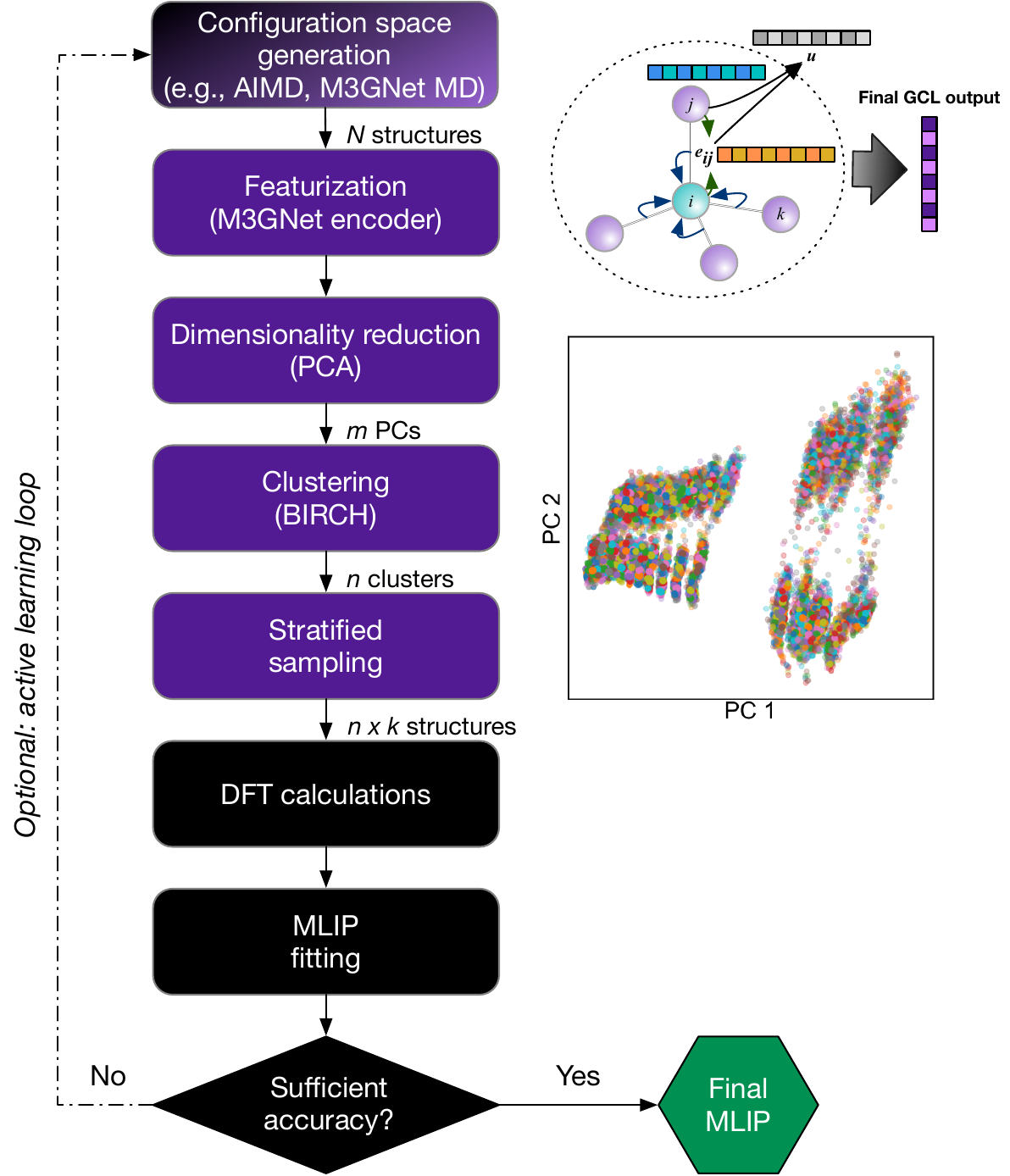}
    \caption{Workflow of DImensionality REduction - Clustering - sTratified (DIRECT) sampling. The standard steps in MLIP development are in black boxes, while the key conceptual improvements proposed in this work are highlighted in purple boxes. The methods in the brackets are those used in the present work, though they can be substituted with other similar approaches.}
    \label{fig:workflow}
\end{figure}

Figure \ref{fig:workflow} provides a workflow of the proposed DImensionality-Reduced Encoded Clusters with sTratified (DIRECT) sampling approach, which comprises five main steps:
\begin{enumerate}
    \item \textbf{Configuration space generation.} A comprehensive configuration space of $N$ structures for the system of interest is generated. This can be performed using commonly employed approaches, such as sampling of trajectories from AIMD simulations and generating structures by applying random atom displacements and lattice strains, or alternatively, by sampling MD trajectories with universal MLIPs, such as M3GNet as demonstrated in later sections. 
    \item \textbf{Featurization/encoding.} Next, the configuration space is featurized into fixed length vectors for each structure. While there are many well-established descriptors used in MLIPs, most describe only the local atomic environments and do not efficiently handle arbitrary chemical complexity. Taking inspiration from the AtomSets framework,\cite{chenAtomSetsHierarchicalTransfer2021} we propose to use the concatenated output of the final graph convolutional layer (GCL) from pre-trained graph deep learning formation energy models that cover diverse chemistries. The rationale for this choice is that the final output layer of such models already encodes a fixed-length structure/chemistry representation for predicting energy. Furthermore, the formation energy is one of the most readily available large datasets in materials databases such as the Materials Project. In this work, we will use the 128-element vector outputs from the M3GNet model trained on the formation energies of materials in Materials Project,\cite{chenUniversalGraphDeep2022} though similar results are obtained with the 96-element vector outputs from the MEGNet formation energy model.\cite{chenGraphNetworksUniversal2019, chenAtomSetsHierarchicalTransfer2021} 
    \item \textbf{Dimensionality reduction.} A further dimensionality reduction step is carried out. Here, we apply principal component analysis (PCA) on the normalized fixed-length features from the encoding step. Following Kaiser's rule, the first $m$ PCs with eigenvalues over 1, i.e., explaining more variance than any single variable, are kept to represent the feature space. For ease of visualization, we have plotted only the first two PCs in Figure \ref{fig:workflow}, even though more than two PCs are usually used in this work.
    \item \textbf{Clustering.} Next, clustering is carried out to group structures with shared characteristics. In this work, the balanced iterative reducing and clustering using hierarchies (BIRCH) algorithm,\cite{zhangBIRCHEfficientData1996} a highly efficient centroid-based clustering method, is used to divide all features into clusters based on their locations in the $m$-D feature space. PCs are weighted by their respective explained variance before clustering. The choice of the number of clusters ($1 \leq n \leq N$) can be determined based on the desired accuracy and computational budget. Figure \ref{fig:workflow} shows the clustering of 50,050 Ti-H M3GNet MD snapshots into 3,000 clusters.
    \item \textbf{Stratified sampling.} Finally, stratified sampling of $k$ structures from each cluster is then performed to construct a robust training set. If $k = 1$, features with the shortest Euclidean distance to the centroid of each cluster will be selected. If $k$ is greater than 1, features in each cluster will be sorted according to their Euclidean distances to the respective centroid, and then $k$ features will be selected at constant index intervals. When $k$ is greater than the size of certain clusters, all data in those clusters will be selected, and the user can choose whether or not to allow duplicated selection. Similarly, the choice of $k$ depends on the desired coverage and computational budget.
\end{enumerate}

The remaining steps are similar to standard MLIP development procedures in the literature. Static DFT calculations are performed on the $M \le n \times k$ structures from the DIRECT sampling procedure. While the DIRECT sampling approach can be integrated into an active learning loop, it is designed to obtain comprehensive coverage of the configuration space of interest \textit{prior to DFT calculations and minimize/eliminate the need for active learning iterations}. It should also be noted that most of the steps in the conceptual DIRECT sampling workflow can be replaced by alternative methods, e.g., the choice of the structure featurizer, the dimensionality reduction technique, and the clustering algorithm.

\subsection{Generating a more diverse Materials Project training set}

\begin{figure}[!htpb]
\centering
\begin{subfigure}[b]{0.31\textwidth}
    \centering
	\includegraphics[width=\textwidth]{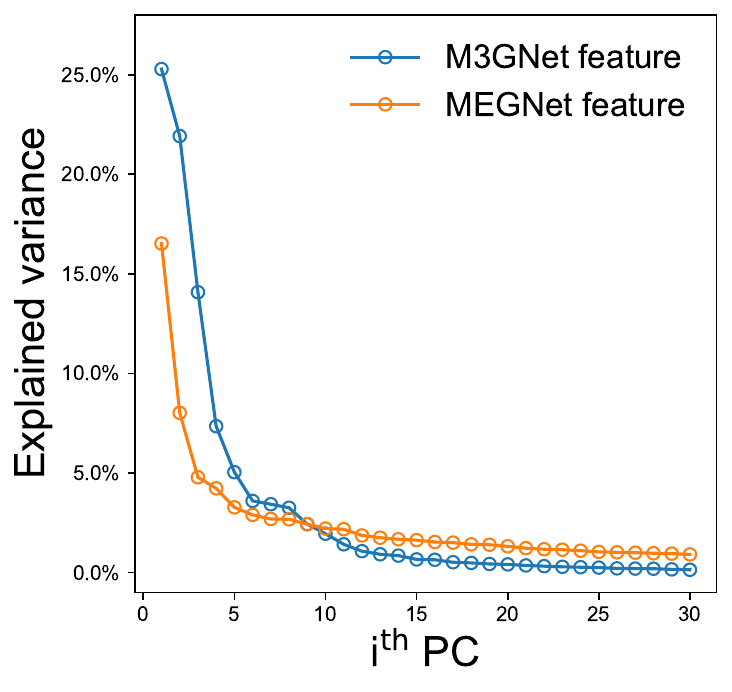}
	\caption{\label{subfig:MP_PCA_explained_variance}}
\end{subfigure}
\begin{subfigure}[b]{0.3\textwidth}
    \centering
	\includegraphics[width=\textwidth]{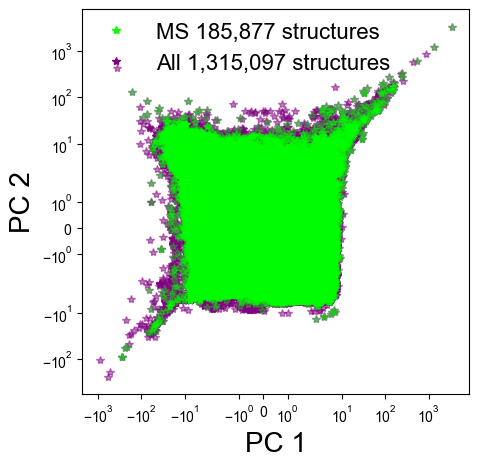}
	\caption{\label{subfig:MP_PCA_MS}}
\end{subfigure}
\begin{subfigure}[b]{0.3\textwidth}
    \centering
	\includegraphics[width=\textwidth]{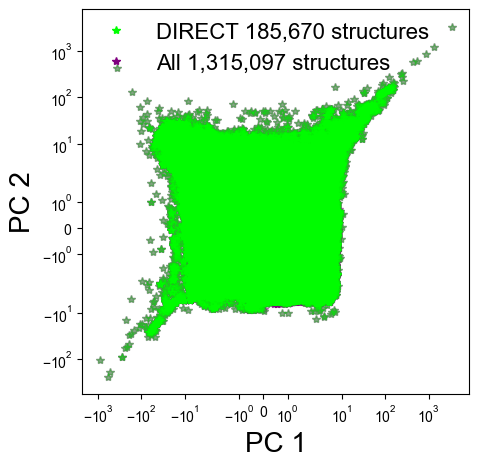}
	\caption{\label{subfig:MP_PCA_CS}}
\end{subfigure}
\begin{subfigure}[b]{\textwidth}
    \centering
	\includegraphics[width=0.95\textwidth]{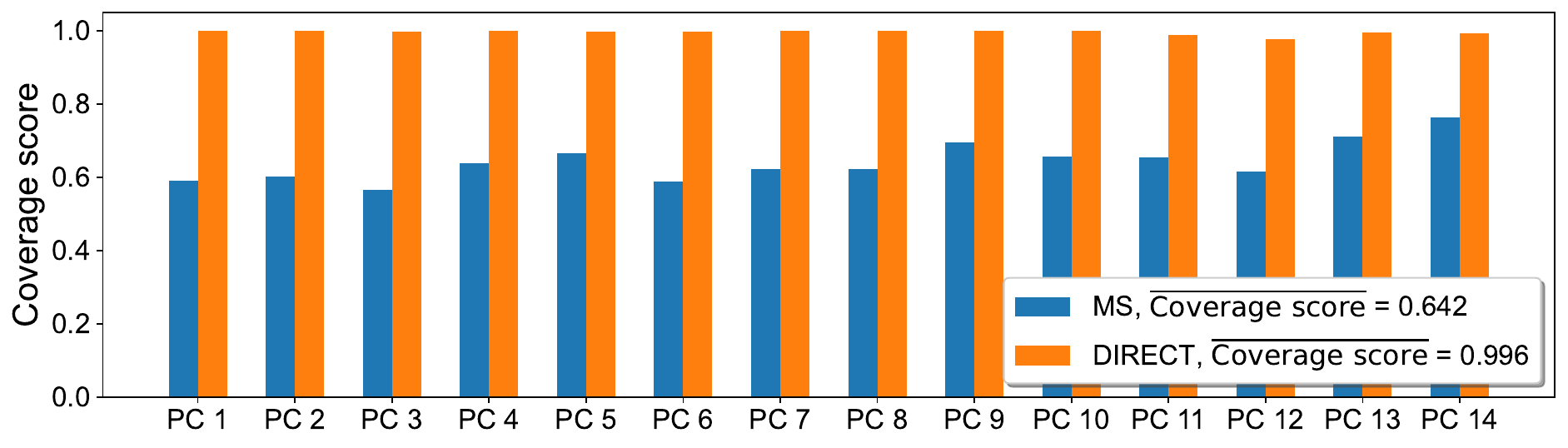}
	\caption{\label{subfig:MP_PCA_coverage_score}}
\end{subfigure}

\caption{\label{fig:MP_PCA_feature_coverage}Comparison of DIRECT versus manual sampling (MS). (a) Explained variance of the first 30 principal components of the encoded features using the M3GNet and MEGNet formation energy models. Visualization of the coverage of the first two PCs of the M3GNet-encoded structure features by (b) MS and (c) DIRECT sets. (d) Feature coverage scores for the first 14 PCs of the M3GNet-encoded structure features by the MS set and DIRECT set.}
\end{figure}

We will first demonstrate the utility of the DIRECT sampling approach using one of the largest datasets that have been used in MLIP fitting - the \textit{MPF.2021.2.8.All} dataset. The \textit{MPF.2021.2.8.All} dataset includes all ionic steps from both the first and second relaxation calculations in Materials Project\cite{jainCommentaryMaterialsProject2013} (see Methods section for details). 

Figure \ref{fig:MP_PCA_feature_coverage} compares the coverage of feature space by manual sampling (MS) and DIRECT sampling approaches on the \textit{MPF.2021.2.8.All} dataset. The MS set, which contains 185,877 structures, is constructed following the approach outlined by \citet{chenUniversalGraphDeep2022}, which selects the first and middle ionic steps of the first relaxation and the last step of the second relaxation. Using DIRECT sampling with $n=20,044$ and $k=20$, i.e., sampling at most 20 structures from each of the 20,044 clusters, a dataset of 185,670 structures is constructed, approximately the same size as the MS set. 

Figure \ref{subfig:MP_PCA_explained_variance} compares the explained variance vs the PCs of the encoded features using the M3GNet and MEGNet formation energy models. It can be seen that the M3GNet-encoded features are significantly more efficient, with a cumulative explained variance of 49\% and 93\% for the first 2 and 14 PCs, respectively. In contrast, the cumulative explained variance for the first 2 and 14 PCs for the MEGNet-encoded features are 25\% and 57\%, respectively. This indicates that the incorporation of the 3-body interactions in M3GNet leads to a more robust encoding of the diverse structures and chemistries in the \textit{MPF.2021.2.8.All} dataset. From the plots of the first two PCs of the M3GNet-encoded features of the MS set (Figure \ref{subfig:MP_PCA_MS}) and DIRECT set (Figure \ref{subfig:MP_PCA_CS}), it can clearly be observed that the MS set undersamples structures located at the boundaries of the feature space, while the DIRECT set provides more comprehensive coverage. The coverage score for the first 14 PCs was calculated as $\sum_{i=1}^{n_b}c_i/n_b$, where the entire range of values for each PC is divided into $n_b$ bins and $c_i$ equals 1 if data in the $i^{th}$ bin is successfully sampled, and 0 otherwise. The coverage score of the entire \textit{MPF.2021.2.8.All} set is 1 by definition. Using $n_b=50,000$, we find that the coverage scores of the DIRECT set across the first 14 PCs are all close to 1, with an average of 0.996, while the coverage scores of the MS set are all below 0.8 with an average of 0.642. Similar trends are observed for the 128-element M3GNet feature space (see Figure S1).

\begin{figure}[H]
\centering
\begin{subfigure}[b]{0.31\textwidth}
    \centering
	\includegraphics[width=\textwidth]{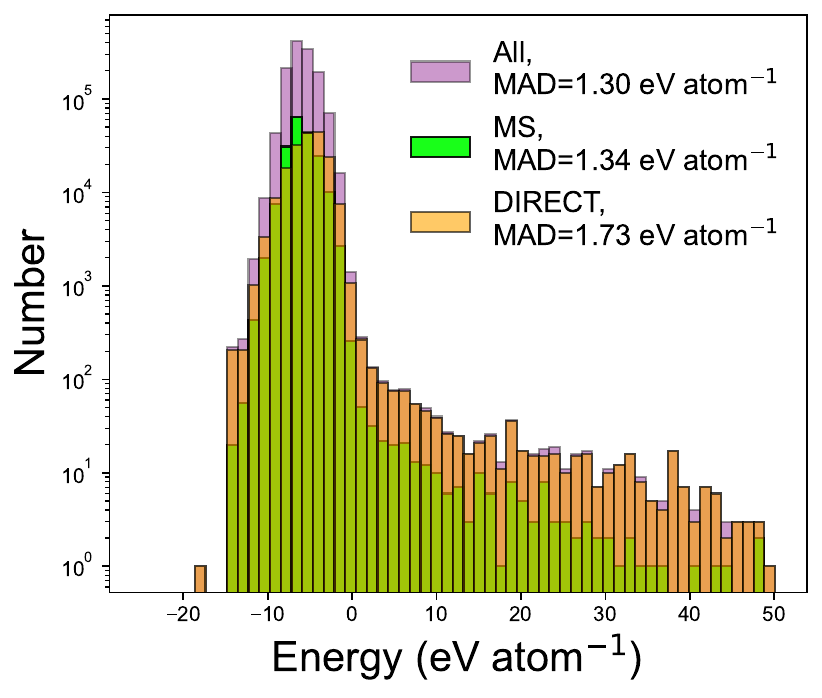}
	\caption{\label{subfig:MP_distribution_energy}}
\end{subfigure}
\begin{subfigure}[b]{0.31\textwidth}
    \centering
	\includegraphics[width=\textwidth]{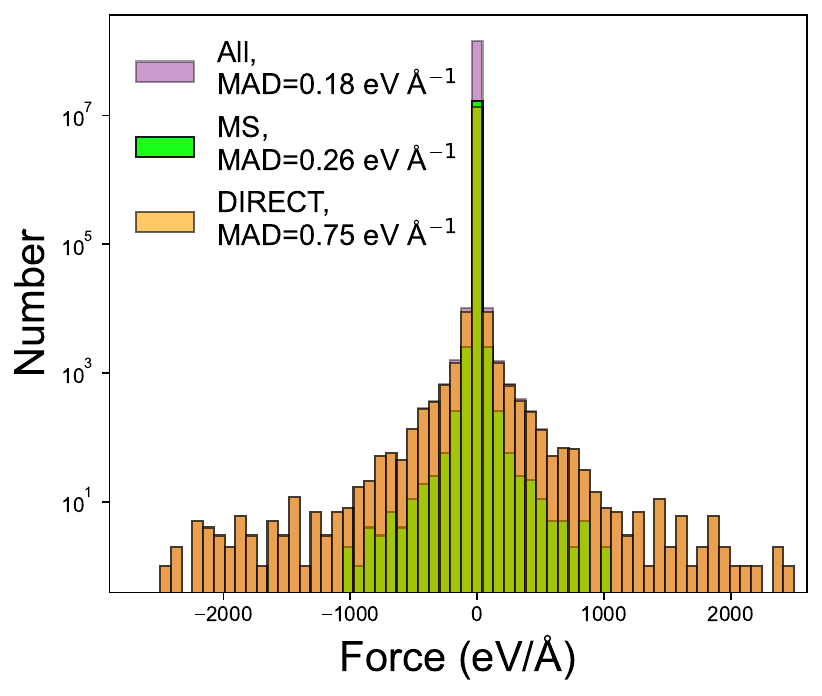}
	\caption{\label{subfig:MP_distribution_force}}
\end{subfigure}
\begin{subfigure}[b]{0.31\textwidth}
    \centering
	\includegraphics[width=\textwidth]{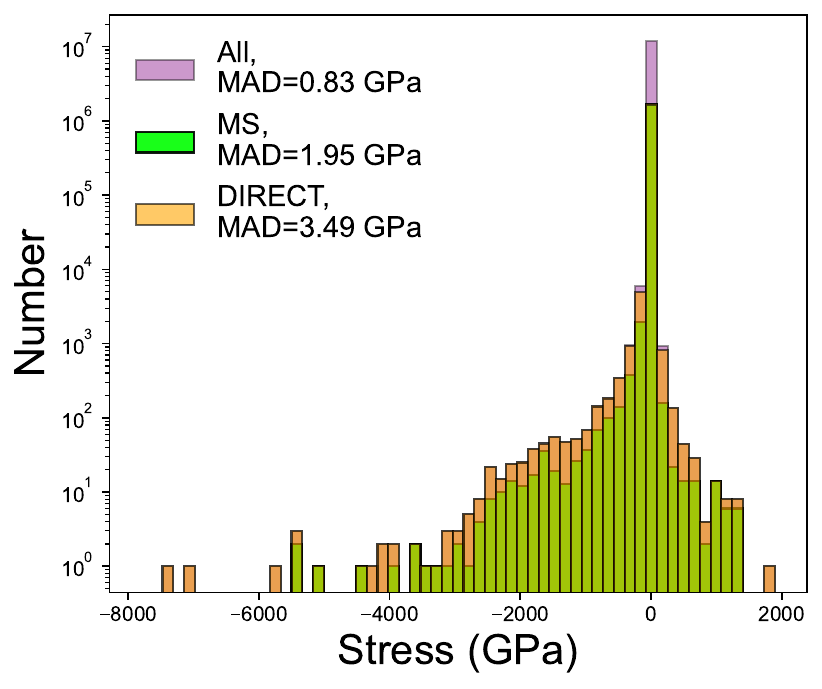}
	\caption{\label{subfig:MP_distribution_stress}}
\end{subfigure}
\caption{\label{fig:MP_EFS_coverage} Distribution of (a) energies, (b) forces and (c) stresses in the DIRECT set, the MS set, and \textit{MPF.2021.2.8.All} (referred as ``All'') are labeled by colors of yellow, green and purple, respectively. Mean absolute deviation (MAD) of each data set is annotated.}
\end{figure}

Figure \ref{subfig:MP_distribution_energy} to \ref{subfig:MP_distribution_stress} compares the distribution of the energies, forces and stresses in the DIRECT and MS sets relative to the entire \textit{MPF.2021.2.8.All} (``All'') dataset. Despite having a comparable total number of structures, the DIRECT set provides a better coverage of the entire configuration space, with a much larger MAD in energies, forces and stresses compared to the MS set. This can be attributed to the better sampling of uncommon local environments in feature space by DIRECT sampling compared to manual sampling.

\subsection{Training a more reliable M3GNet universal potential}

\begin{figure}[H]
\centering
\begin{subfigure}[b]{0.31\textwidth}
    \centering
	\includegraphics[width=\textwidth]{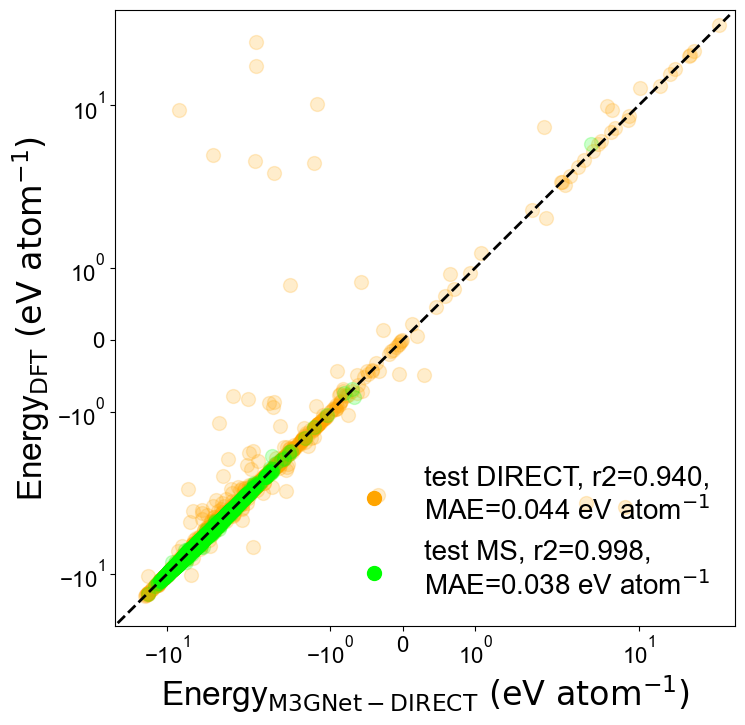}
	\caption{\label{subfig:parity_M3GNet_DIRECT_E}}
\end{subfigure}
\begin{subfigure}[b]{0.31\textwidth}
    \centering
	\includegraphics[width=\textwidth]{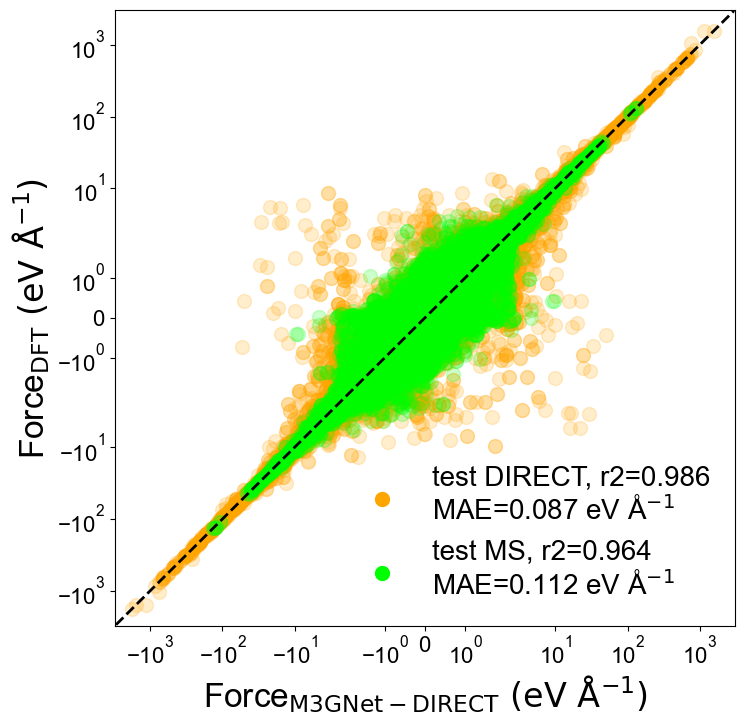}
	\caption{\label{subfig:parity_M3GNet_DIRECT_F}}
\end{subfigure}
\begin{subfigure}[b]{0.31\textwidth}
    \centering
	\includegraphics[width=\textwidth]{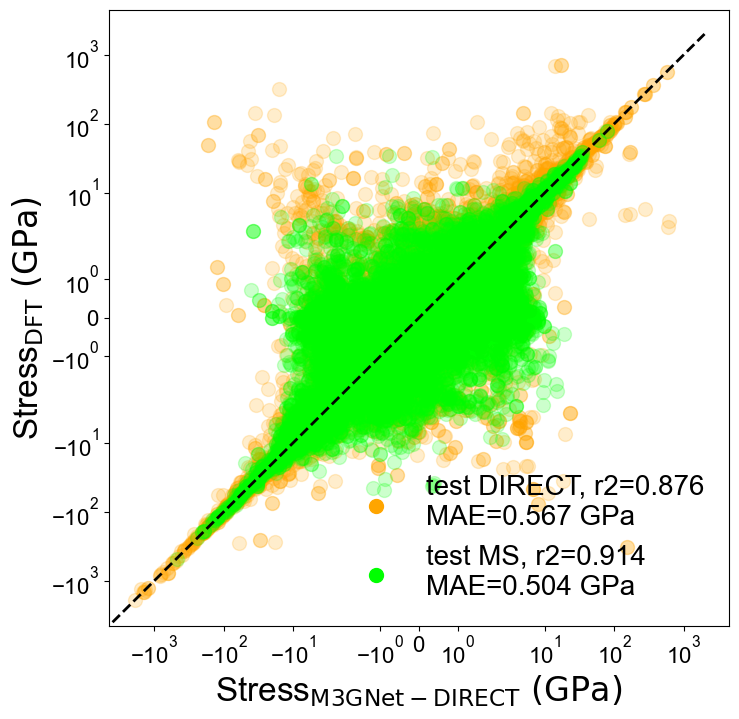}
	\caption{\label{subfig:parity_M3GNet_DIRECT_S}}
\end{subfigure}
\begin{subfigure}[b]{0.31\textwidth}
    \centering
	\includegraphics[width=\textwidth]{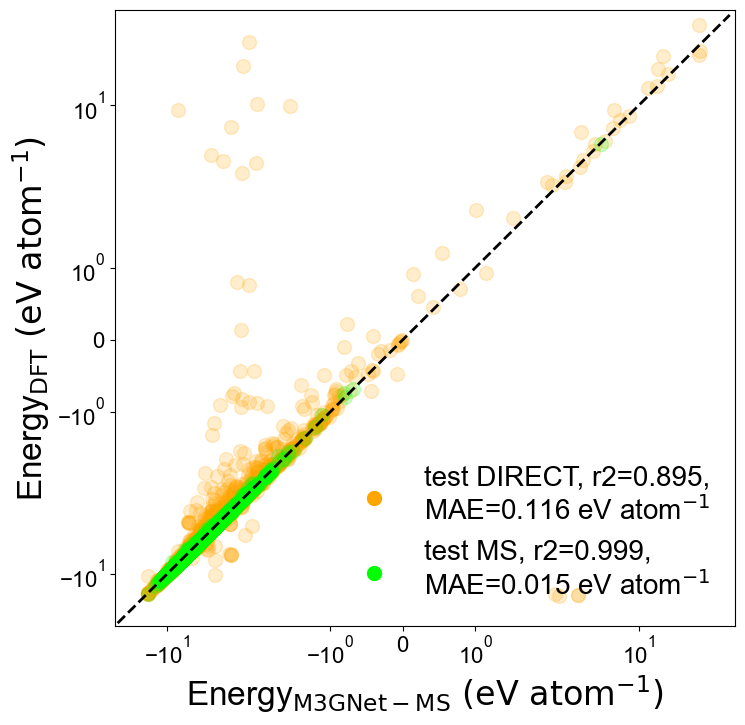}
	\caption{\label{subfig:parity_M3GNet_MS_E}}
\end{subfigure}
\begin{subfigure}[b]{0.31\textwidth}
    \centering
	\includegraphics[width=\textwidth]{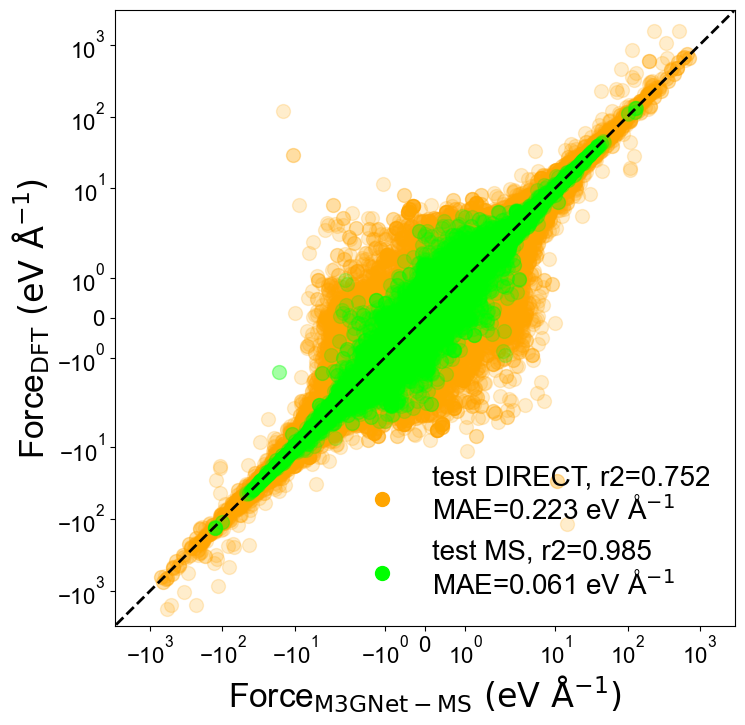}
	\caption{\label{subfig:parity_M3GNet_MS_F}}
\end{subfigure}
\begin{subfigure}[b]{0.31\textwidth}
    \centering
	\includegraphics[width=\textwidth]{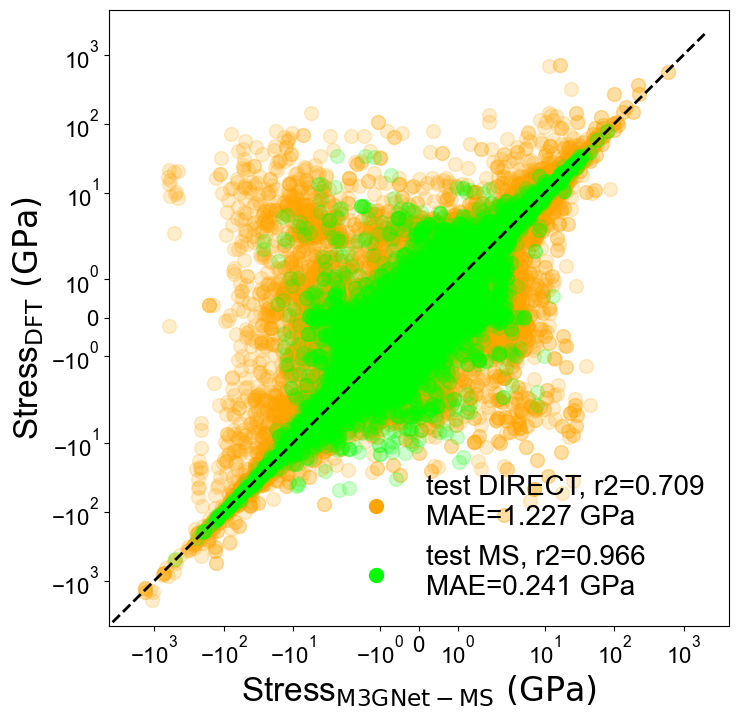}
	\caption{\label{subfig:parity_M3GNet_MS_S}}
\end{subfigure}
\begin{subfigure}[b]{0.31\textwidth}
    \centering
	\includegraphics[width=\textwidth]{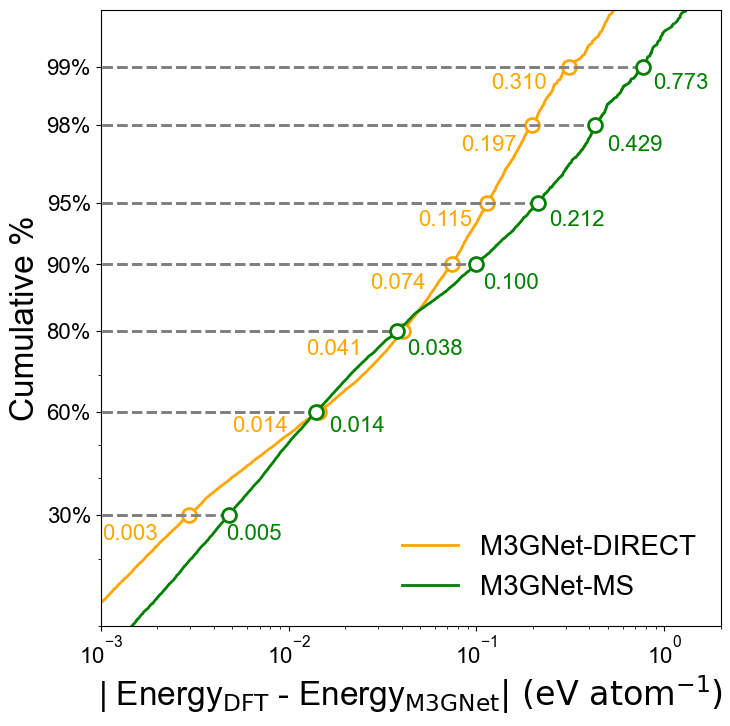}
	\caption{\label{subfig:m3gnet_cumulative_error_MP_E}}
\end{subfigure}
\begin{subfigure}[b]{0.31\textwidth}
    \centering
        \includegraphics[width=\textwidth]{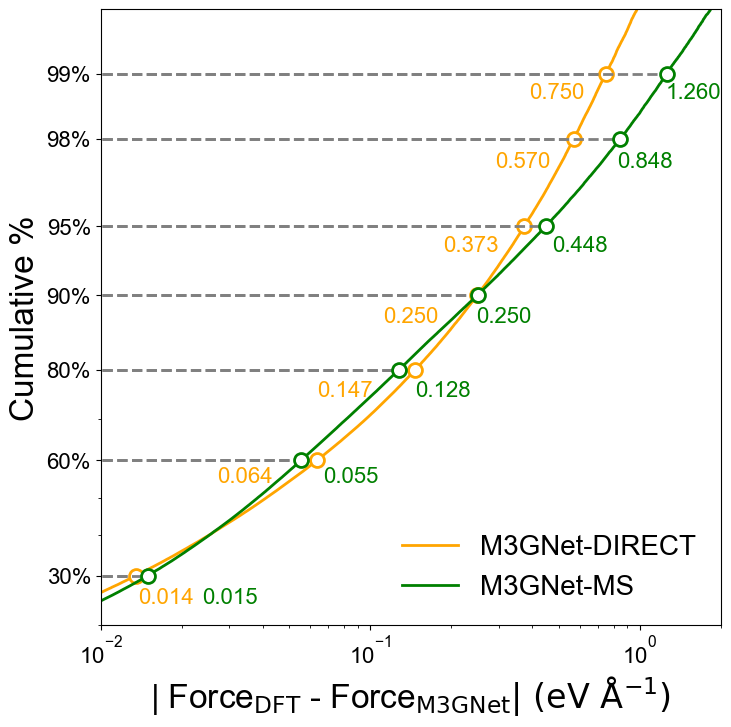}
    \caption{\label{subfig:m3gnet_cumulative_error_MP_F}}
\end{subfigure}
\begin{subfigure}[b]{0.31\textwidth}
    \centering
        \includegraphics[width=\textwidth]{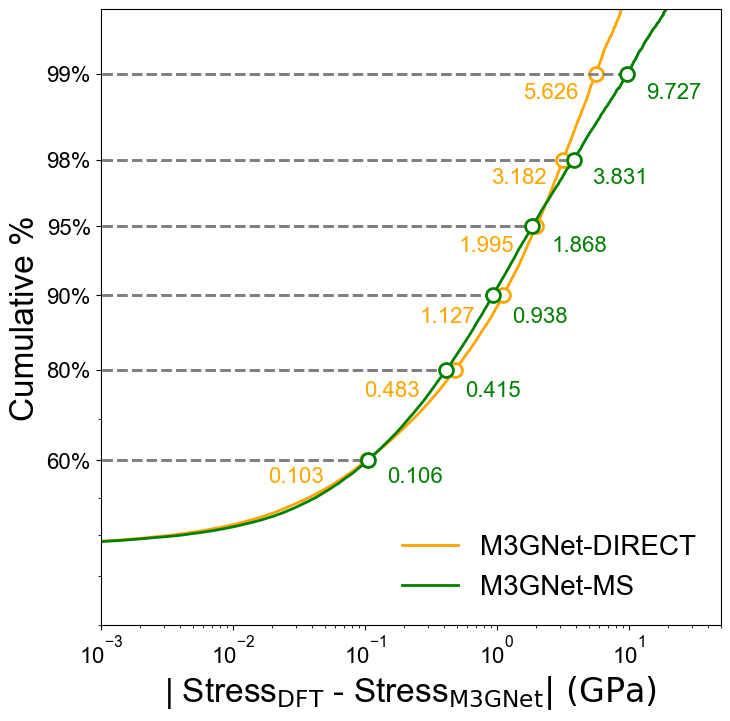}
    \caption{\label{subfig:m3gnet_cumulative_error_MP_S}}
\end{subfigure}

\caption{\label{fig:MP_M3GNet_EFS_parity}Performance of M3GNet universal potentials (UPs) trained using the DIRECT and MS training sets. Parity plots for (a) energies, (b) forces and (c) stresses for the M3GNet UP trained on the DIRECT set. The equivalent plots for the M3GNet UP trained on the MS set is shown in plots (d)-(f). The cumulative errors of (g) energies, (h) forces and (i) stresses in the two test sets by the two UPs are also plotted.}
\end{figure}

Figure \ref{fig:MP_M3GNet_EFS_parity} compares the performance of the M3GNet UPs trained on the DIRECT and MS sets (referred to as M3GNet-DIRECT and M3GNet-MS, respectively) relative to the ground truth DFT. The training protocols are largely similar to the ones used in the original M3GNet UP, with minor modifications as outlined in the Methods section. Two test sets were constructed from a random sample of 5\% of the DIRECT set and MS set, excluding any structures showed up in the training and validation processes of both M3GNet-DIRECT and M3GNet-MS. These two test sets are labelled as ``test DIRECT`` and ``test MS``, containing 7,635 and 7,684 structures, respectively.

As expected, both M3GNet UPs provide the best performance on their respective datasets, i.e., the M3GNet-DIRECT outperforms M3GNet-MS on the DIRECT test set, while the reverse is true on the MS test set (Figure \ref{fig:MP_M3GNet_EFS_parity}a-f). However, the M3GNet-MS UP performs significantly worse on the DIRECT test set than on the MS test set, with MAEs in energies and forces that are about an order of magnitude larger than those for the MS test set. In contrast, the M3GNet-DIRECT UP has comparable MAEs across both the DIRECT and MS test sets. It should be stressed that the DIRECT test set is the more challenging of the two, with a greater number of data points with large energies, forces, and stresses as well as large MADs. The average MAEs in energies, forces, and stresses of the M3GNet-DIRECT UP across both the DIRECT and MS test sets (0.041 eV atom$^{-1}$, 0.101 eV \AA$^{-1}$ and 0.54 GPa, respectively) are only slightly higher than the test errors reported for the original M3GNet UP (0.035 eV atom$^{-1}$, 0.072 eV Å$^{−1}$ and 0.41 GPa, respectively.\cite{chenUniversalGraphDeep2022} Figure \ref{fig:MP_M3GNet_EFS_parity}g-i shows the cumulative error distribution for the M3Gnet UPs on the combined data in ``test DIRECT`` and ``test MS``. While M3GNet-MS and M3GNet-DIRECT have relatively have similar errors for $\sim$90\% of the test structures, M3GNet-DIRECT significantly outperforms M3GNet-MS (lower energy, force and stress errors) for the remaining $\sim$10\% of test structures that are most challenging.

\begin{figure}[H]
    \centering
\begin{subfigure}[b]{0.45\textwidth}
    \centering
    \includegraphics[width=\textwidth]{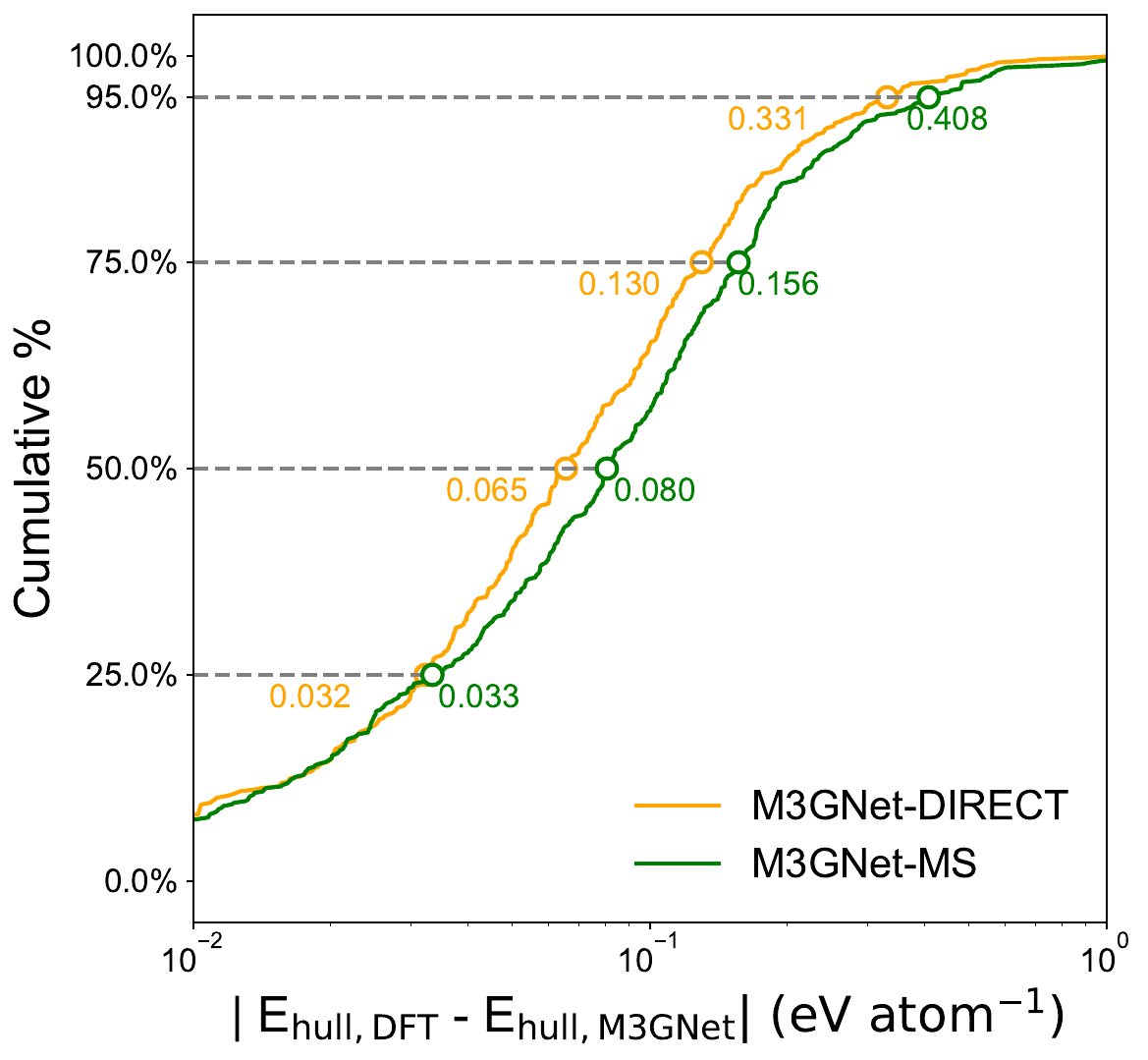}
    \caption{}\label{subfig:hypo_O-containing}
\end{subfigure}    
\begin{subfigure}[b]{0.45\textwidth}
    \centering
    \includegraphics[width=\textwidth]{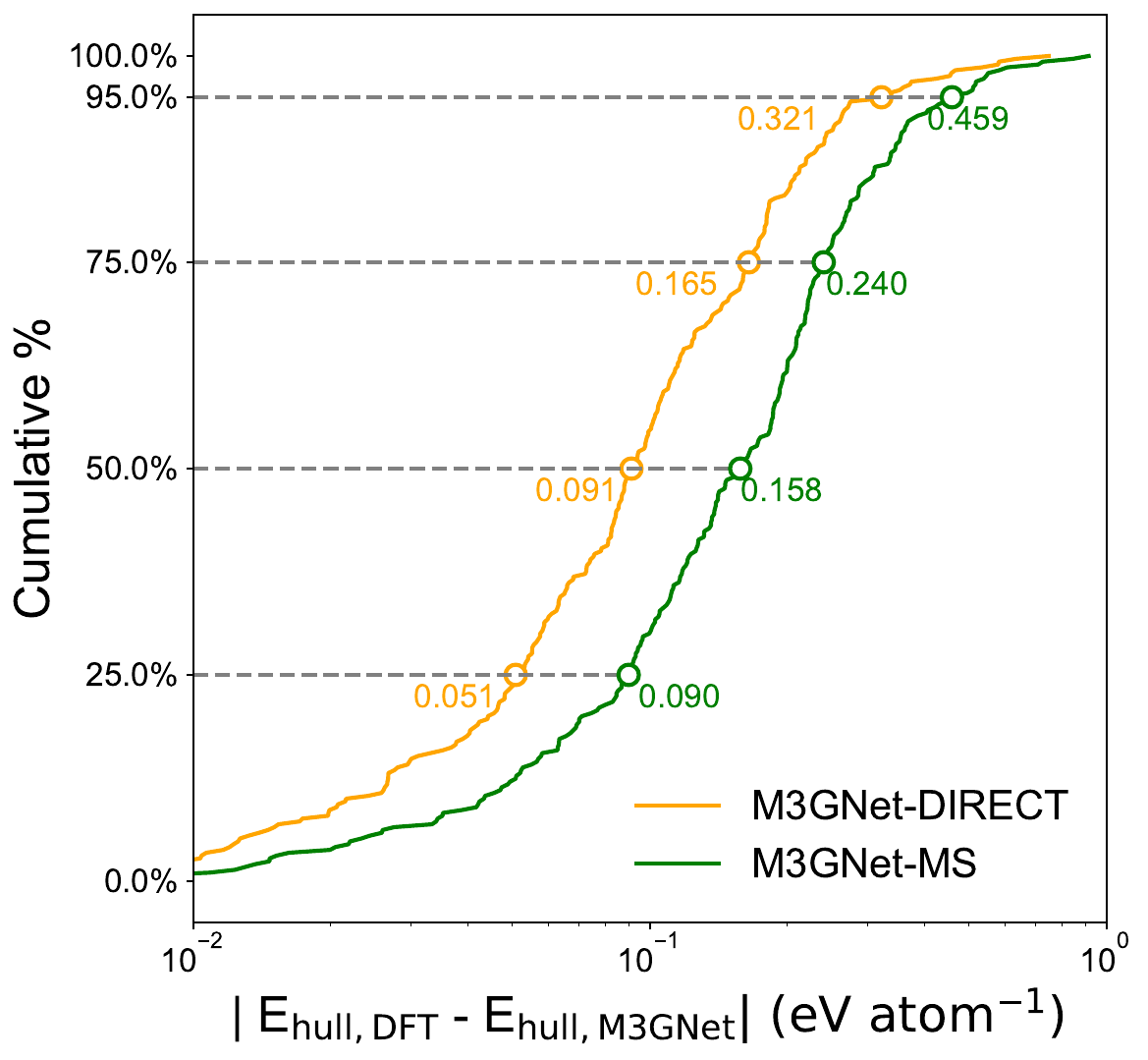}
    \caption{}\label{subfig:hypo_S-containing}
\end{subfigure}
\caption{\label{fig:MP_M3GNet_hypo_ehull_extrapolation}Cumulative absolute errors for energy above hull $E_{hull}$ prediction for (a) O- and (b) S-containing hypothetical materials by M3GNet-DIRECT and M3GNet-MS UPs.}
\end{figure}

To further assess the performance of the M3GNet UPs on unseen structures, we have calculated the energies above hulls for a dataset of 506 O-containing compounds, 291 S-containing hypothetical materials. This dataset was randomly selected from the $\sim$30 million hypothetical materials generated by \citet{chenUniversalGraphDeep2022} (see Methods for details).  

From Figure \ref{fig:MP_M3GNet_hypo_ehull_extrapolation}, the M3GNet-DIRECT UP provides an improved prediction of $E_{hull}$ for hypothetical materials compared to the M3GNet-MS UP. While the performance improvement is relatively small for the O-containing compounds, a significant reduction in error is observed for the S-containing compounds.  For instance, the error in $E_{hull}$ by M3GNet-DIRECT and M3GNet-MS is below 0.091 and 0.158 eV atom$^{-1}$, respectively for 50\% of the S-containing hypothetical materials. The distribution of errors for the M3GNet-DIRECT is comparable between the O-containing and S-containing hypothetical materials. In contrast, the M3GNet-MS UP performs much worse for the S-containing materials relative to the O-containing materials. These observations can be attributed to the fact that the original Materials Project dataset contains a preponderance of O-containing materials (34,556) and significantly fewer S-containing materials (5,333) among the 62,783 compounds.

\subsection{Developing an accurate MLIP for titanium hydrides}
As illustrated in Figure~\ref{fig:workflow}, the two most computationally intensive steps in the development of MLIPs are the generation of the configuration space and DFT calculations of energies and forces. Often-used strategies to sample configuration space include highly expensive AIMD simulations and iterative efforts in active learning (AL) workflows. The advent of UPs such as M3GNet can provide the means to bypass \textit{ab initio} methods and minimize or even eliminate AL iterations for the generation of a diverse configuration space. 

In this section, we demonstrate the capability of the DIRECT sampling approach combined with the M3GNet UP to construct reliable MLIPs. Here we have chosen the moment tensor potential (MTP) to study titanium hydrides (TiH$_n$), which are promising materials for hydrogen storage.~\cite{zhu2022} Hydrogen is well-known to be highly diffusive in these systems, even at ambient temperatures, and a relatively short time step, e.g., $\sim 0.5$ fs, is required for stable MD simulations. Therefore, this system provides a robust test for our proposed workflow. Moreover, we note that these descriptor-based MLIPs are more computationally efficient to study particular chemistry due to their low model complexity relative to UPs.

\subsubsection{Configuration space for Ti-H}

\begin{figure}[H]
    \centering
    \includegraphics[width=0.5\textwidth]{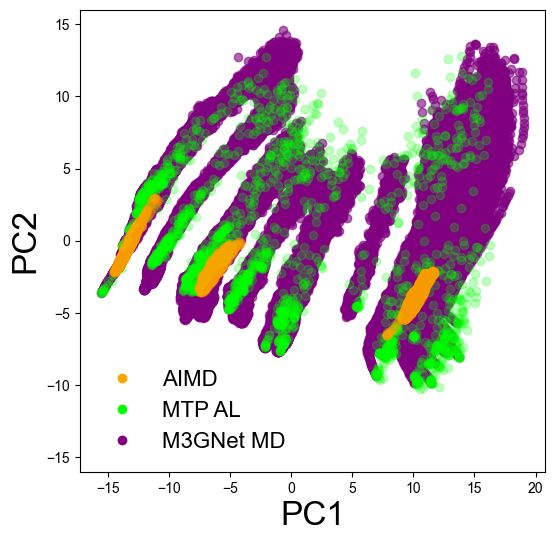}
\caption{\label{fig:TiH_configuration_space}Plot of the first two principal components of the feature space of \ce{TiH_n} ($0\le n \le 2$) sampled by structures from three different sources, i.e., 75,000 AIMD snapshots for HCP \ce{Ti36H2}, BCC \ce{Ti36H36} and FCC \ce{Ti36H72} at 1000 K, 2,063 configurations from MTP AL and 273,000 MD NpT snapshots from M3GNet-DIRECT. \ce{H2} structures are excluded in this analysis to ensure better resolution for Ti-containing structures.}
\end{figure}      
 
To generate a comprehensive configuration space for \ce{TiH_$n$}, multi-temperature MD simulations using the M3GNet-DIRECT UP were carried out on a comprehensive set of hydrogen compositions (see Methods section). For comparison, two other structure sets were constructed by AIMD and MTP AL, including: (i) 75,000 snapshots of AIMD simulations of HCP \ce{Ti36H2}, BCC \ce{Ti36H36} and FCC \ce{Ti36H72} at 1000 K; and (ii) 2,077 structures collected by MTP AL for the same 274 MD scenarios applied by M3GNet MD. Figure \ref{fig:TiH_configuration_space} reports the first two principal components (PCs) of M3GNet-encoded features for each structure generation method. It is shown that AIMD simulations only sample a small part of the configuration space due to the short simulation time scales and limited structure diversity. In contrast, MD simulations using the M3GNet-DIRECT UP sample the largest configuration, encompassing most of the configuration space visited by the MTP AL and AIMD simulations. Coverage of the few structures visited by the AL process that lies outside of the M3GNet-DIRECT set may require additional MD conditions and/or the coupling of AL with our proposed DIRECT sampling strategy.

\subsubsection{MTP fitting}

\begin{figure}[H]
\centering
\begin{subfigure}[b]{0.45\textwidth}
    \centering
	\includegraphics[width=\textwidth]{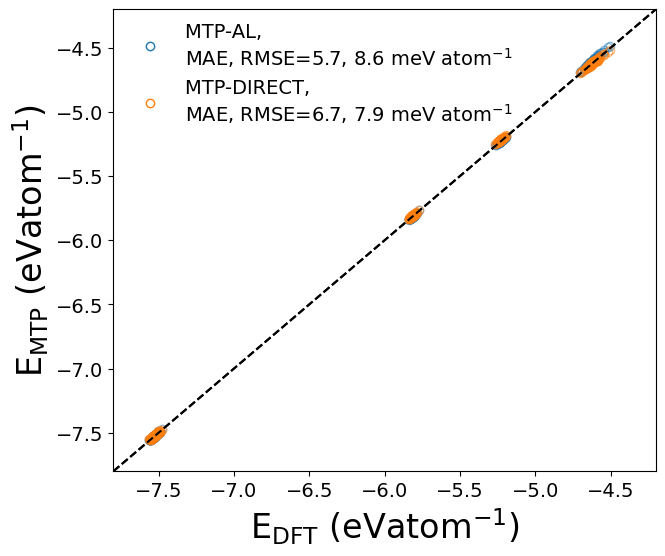}
	\caption{\label{subfig:TiH_parity_E}}
\end{subfigure}
\begin{subfigure}[b]{0.45\textwidth}
    \centering
	\includegraphics[width=\textwidth]{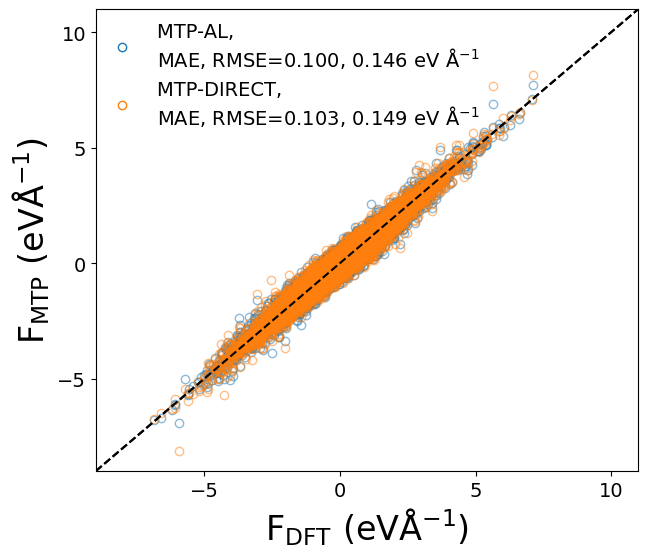}
	\caption{\label{subfig:TiH_energy_parity_F}}
\end{subfigure}
\caption{\label{fig:TiH_parity_EFS}Energy and force errors of 400 AIMD test structures by MTPs fitted using M3GNet-DIRECT MD structures (MTP-DIRECT) and MTP AL structures (MTP-AL).}
\end{figure}

A DIRECT sampling is applied to select 1 structure each from 954 clusters of the 274,000 M3GNet MD snapshots. DFT static calculations of 947 successfully converged and were used as the training set for MTP-DIRECT (see details in Methods). To compare the accuracy of energy and force predictions by "MTP-DIRECT" and "MTP-AL", 400 AIMD snapshots are selected as test structures from four 10-ps AIMD trajectories (HCP \ce{Ti36H2} at 1000 K, BCC \ce{Ti36H36} at 1000 K and FCC \ce{Ti36H72} at 1000 and 3000 K) at 0.1 ps intervals. DFT static calculations were then performed to obtain energies and forces as test data. As shown in Figure \ref{fig:TiH_parity_EFS}, both MTPs provide comparable test MAE and RMSE below 10 eV atom$^{-1}$ and 0.15 eV \AA$^{-1}$~for energies and forces, respectively. More importantly, predictions by both MTPs lie closely to the diagonal line, indicating their high reliability for the predictions of Ti-H energies and forces. However, it should be noted that the size of the MTP-DIRECT training set (947) is half of that generated by the MTP AL process (2,077).

\begin{figure}[H]
    \centering
    \includegraphics[width=0.7\textwidth]
    {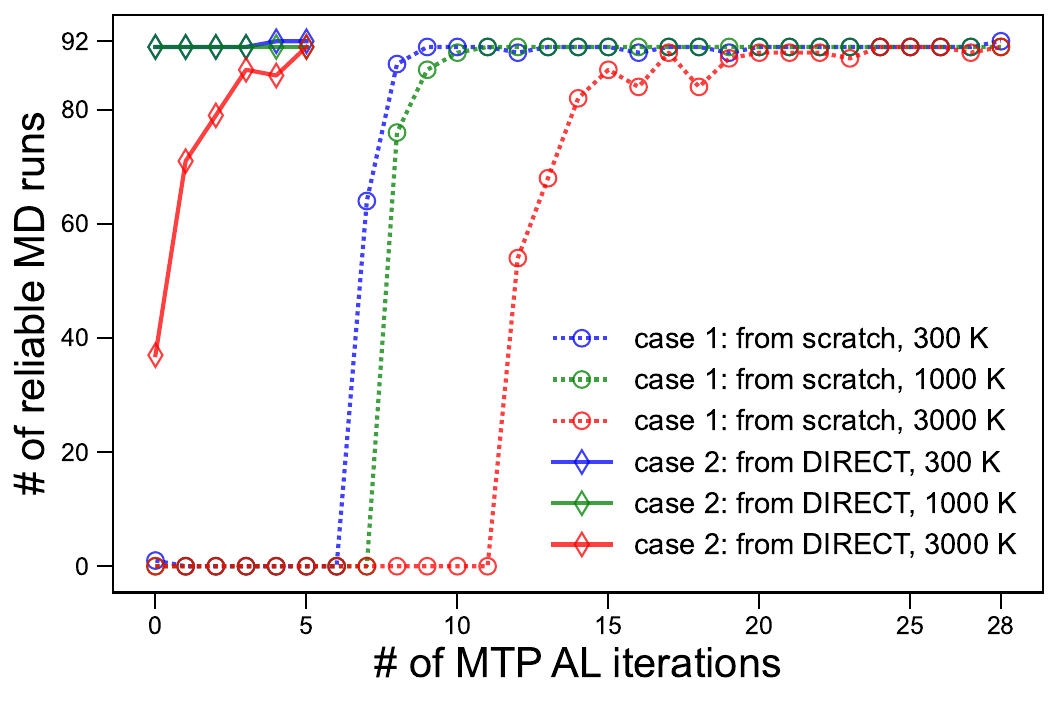}
    \caption{The evolution of MTP MD stability by AL starting from two different initial training sets, i.e., 947 training structures for MTP-DIRECT (labeled as "case 1") and 92 starting structures for the 274 AL scenarios (labeled as "case 2"). Evolution at the three different AL temperatures are plotted separately. }
    \label{fig:TiH_AL_evolution}
\end{figure}

To further evaluate the MD reliability of MTP-DIRECT, the same AL process used to train MTP-AL is applied to MTP-DIRECT. One MD run is considered reliable if no snapshots throughout the 10 ps are having extrapolation grade ($\gamma$) over 3, which is a fairly strict threshold.\cite{gubaevAcceleratingHighthroughputSearches2019} As shown in Figure \ref{fig:TiH_AL_evolution}, all MD runs by MTP-DIRECT at 300 and 1000 K for the 91 Ti-containing Ti-H structures successfully completed at the 0$^{th}$ AL iteration without emergence of any structures with $\gamma$ over 3, indicating that the MTP-DIRECT is already reliable without AL optimization for those MD scenarios. After just 5 AL iterations, MTP-DIRECT can reliably complete the 91 MD runs for Ti-H at 3000 K and the MD run for \ce{H72} at 300 K. In comparison, AL from scratch, i.e., using the 92 MD starting structures as initial training set, took $\sim$ 10 and $\sim$ 20 AL iterations to reach full reliability of MD runs for Ti-H structures at 300 K and at 3000 K, respectively. It took 28 iterations to reliably complete the MD simulation of \ce{H72} at 300 K. Hence, the application of DIRECT sampling to construct the initial MTP-DIRECT reduces the number of AL iterations to reach MD reliability by 75\% and the number of static DFT calculations by 50\% in comparison to the simple AL scheme.

\subsubsection{Hydrogen diffusivity in titanium}

\begin{figure}[H]
\centering
\begin{subfigure}[b]{0.6\textwidth}
    \centering
	\includegraphics[width=\textwidth]{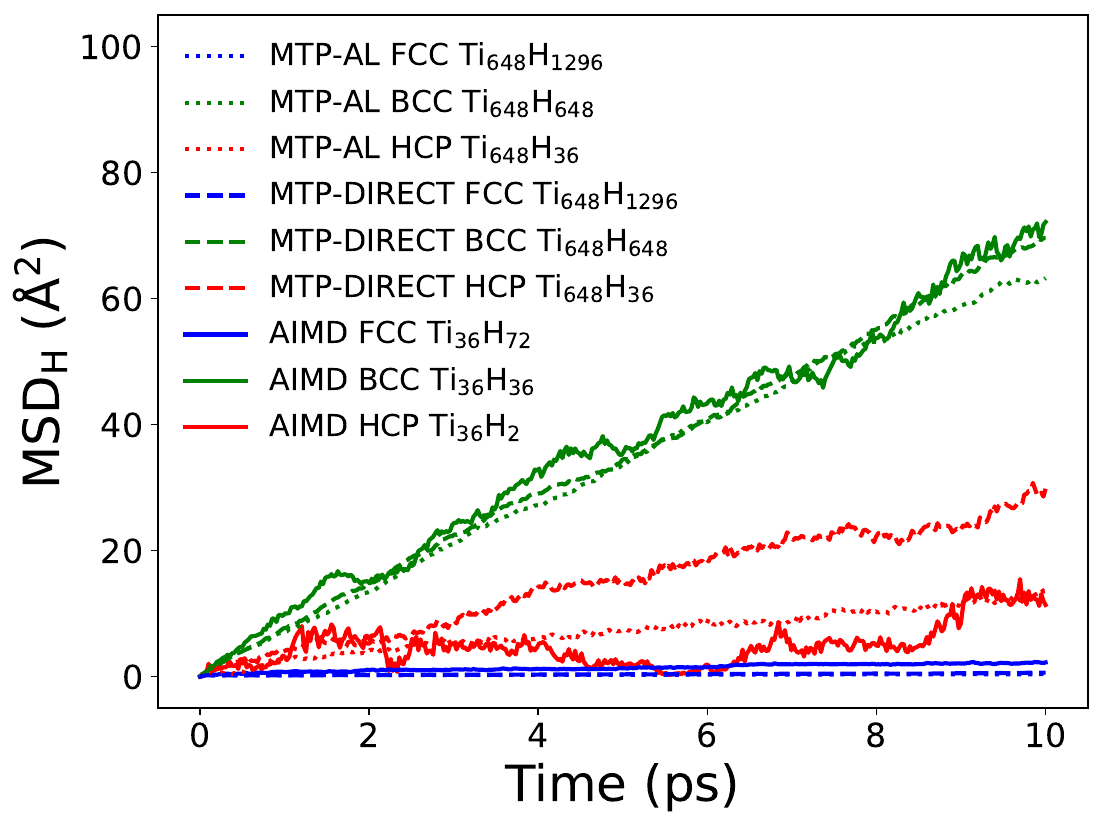}
	\caption{\label{subfig:TiH_MSD_1000K_AIMD_MTPs}}
\end{subfigure}
\begin{subfigure}[b]{0.6\textwidth}
    \centering
	\includegraphics[width=\textwidth]{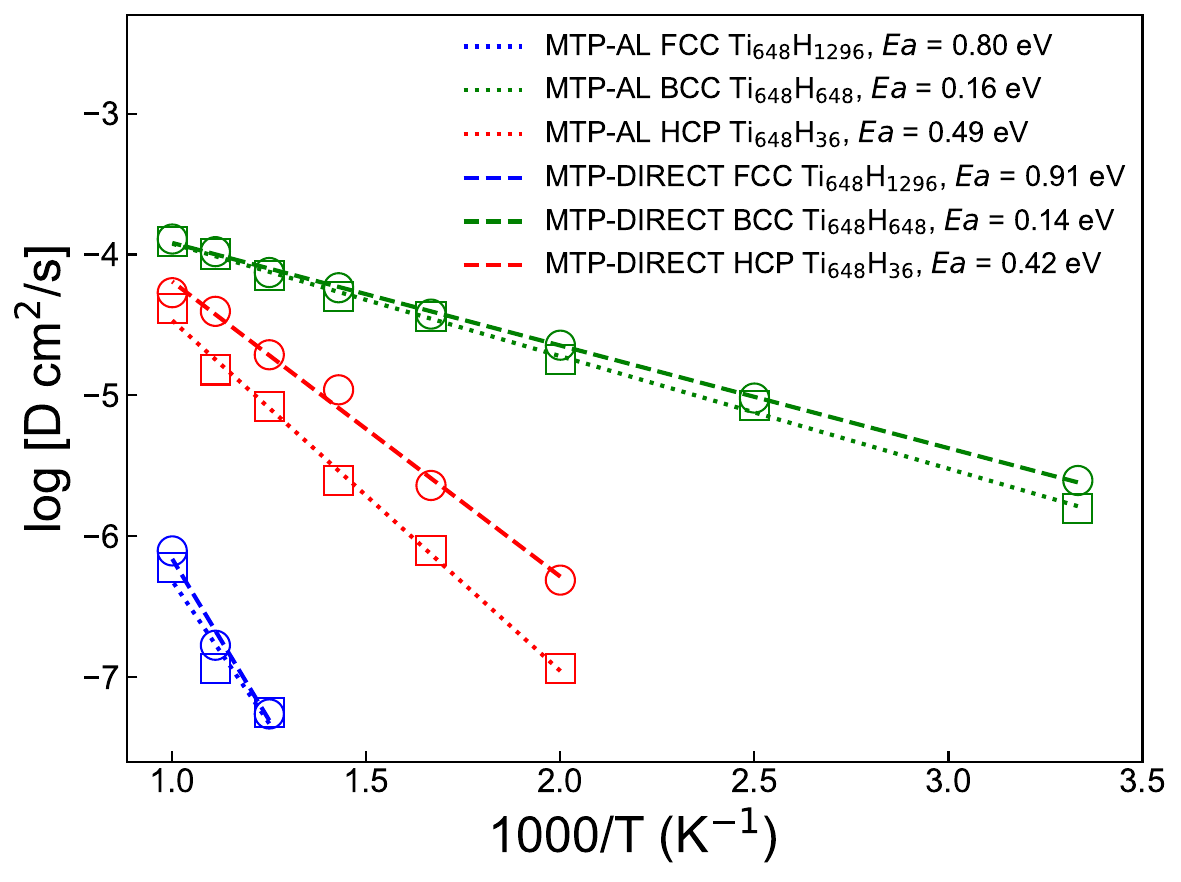}
	\caption{\label{subfig:TiH_Arrhenius_MTPs}}
\end{subfigure}
\caption{\label{fig:TiH_diffusion}MD simulations of H diffusion in HCP \ce{Ti36H2}, BCC \ce{Ti36H36} and FCC \ce{Ti36H72} by AIMD, MTP-AL and MTP-DIRECT. (a) Mean squared displacement (MSD) of H atoms throughout 10-ps AIMD NVT and MTP MD/NVT simulations at 1000 K. (b) Arrhenius plot based on 1-ns MD/NpT simulations by the two MTPs from 300 to 1000 K with 100 K intervals. Diffusivities plotted only above temperatures where sufficient diffusion events are observed for a rigorous analysis. Activation energies ($E_a$) are also indicated in the legend. For MTP MDs, much larger $3\times3\times2$ supercells of the AIMD cells were used.}
\end{figure}

An accurate prediction of hydrogen diffusion in titanium hydrides is challenging because of the nature of H atoms and the complex phase diagram of the systems, comprising HCP, BCC and FCC phases at different hydrogen atomic percentages. To further compare the two MTPs, MD simulations were carried out to investigate hydrogen diffusion in HCP \ce{Ti36H2}, BCC \ce{Ti36H36} and FCC \ce{Ti36H72}. As reported in Figure \ref{subfig:TiH_MSD_1000K_AIMD_MTPs}, both MTPs reproduce the trends of hydrogen MSD at 1000 K predicted by AIMD NVT simulations, i.e., the mean square displacement (MSD) of H is highest in the BCC phase and lowest in the FCC phase. The larger fluctuation of hydrogen MSD in the HCP \ce{Ti36H2} phase in AIMD simulations can be attributed to the limited number of hydrogen atoms in the AIMD cell. This is largely ameliorated with the use of $3\times3\times2$ supercells in MTP simulations, thanks to the computational efficiency of the MLIPs. 

We then carried out 1-ns MD/NpT simulations with MTP-AL and MTP-DIRECT to study hydrogen diffusivity throughout a temperature range of 300-1000 K with a 100 K interval. As shown in the Arrhenius plot in Figure \ref{subfig:TiH_Arrhenius_MTPs}, both MTPs exhibit good agreement on the predicted activation energy ($E_a$) and diffusivity. The simulated $E_a$ of hydrogen diffusion by MTP-DIRECT and MTP-AL are 0.42 and 0.49 eV in the HCP \ce{Ti36H2} phase, 0.14 and 0.16 eV in the BCC \ce{Ti36H36} phase, and 0.91 and 0.80 eV in FCC \ce{Ti36H72} phase, respectively. These results are in excellent agreement with experimentally measured $E_a$ of 0.45 eV in HCP Ti from 873 to 1298 K,\cite{miyoshiDiffusionHydrogenTitanium1996} 0.15 eV in BCC Ti from 555 to 625 K,\cite{sevillaHydrogenDiffusionBcc1988} and 0.92 eV in FCC Ti from 670 to 880 K\cite{kaessHydrogenDeuteriumDiffusion1997}. Meanwhile, the experimentally measured hydrogen diffusivities are $3\times10^{-5}$, $8\times10^{-5}$ and $6\times10^{-8} cm^2/s$ for HCP, BCC and FCC at 1000 K, 800 K and 800 K, respectively, which are in line with the predictions from both MTPs.

\section{Discussion}

In summary, we have demonstrated a robust DImensionality-Reduced Encoded Clusters with sTratified (DIRECT) sampling approach to generate training structures for MLIP development. We also demonstrated that MD simulations using the M3GNet universal potential can be used to generate an initial large configuration space for DIRECT sampling. In many cases, a satisfactory, stable MLIP can be obtained with DIRECT sampling without AL. Even when AL is necessary to further fine-tune the MLIP, DIRECT sampling significantly reduces the number of AL cycles and the total number of DFT static calculations required - the most computationally expensive step in MLIP development.

In this work, we have used the final GCL output vector from a pre-trained M3GNet formation energy model as the structure encoder. We believe this to be a reasonable choice given that the M3GNet formation energy model has been trained on a diverse range of structures and chemistries. The final GCL output, therefore, encodes all relevant chemical information for energy prediction. To our knowledge, there are few other structure encoders that currently satisfy this requirement.

The training cost of the MLIP is controlled by two parameters - the number of clusters $n$ and the number of samples per cluster $k$. For a given computational budget of $M$ DFT static calculations, there can be several choices of $n$ and $k$ for a total configuration space of $N$ structures. As a rule of thumb, one should bias towards having a large number of clusters $n$, i.e., $n \approx M$, to ensure coverage of the extrema of the configuration space. However, $k > 1$ can be used to reduce the CPU and memory requirements for clustering when $n = M$ is not feasible for large $N$ and $M$. This sampling approach also enables an ``interlacing'' approach to building MLIPs. For instance, one can build an initial MLIP using $k = 1$, and increasing $k$ if a higher resolution coverage of the configuration space is deemed necessary for an accurate MLIP.

Finally, we note that DIRECT sampling is agnostic to the chosen MLIP architecture. Here, we have demonstrated its application via the training of an M3GNet universal potential with improved extrapolability and a reliable moment tensor potential for Ti-H. DIRECT sampling can also be used to create datasets with improved structure and chemical diversity to benchmark different MLIP architectures. This work paves the way towards robust development and assessment of MLIPs across any compositional complexity.

\section{Methods}
\subsection{Datasets}
\subsubsection{Materials Project \textit{MPF.2021.2.8.All} Dataset}

The Materials Project dataset used in this work is similar to the \textit{MPF.2021.2.8} dataset used by \citet{chenUniversalGraphDeep2022} in the fitting of the M3GNet UP.\cite{chenUniversalGraphDeep2022, chenchiMPF20212022} The \textit{MPF.2021.2.8} dataset comprises 187,687 ionic steps of 62,783 compounds in the MP database as of Feb 8, 2021.\cite{chenchiMPF20212022, chenUniversalGraphDeep2022} However, while the \textit{MPF.2021.2.8} dataset samples the first and middle ionic steps of the first relaxation and the last step of the second relaxation for calculations in the Materials Project, our initial, unsampled dataset includes all ionic steps from both the first and second relaxation calculations in Materials Project.\cite{jainCommentaryMaterialsProject2013} In addition to the existing filters applied in \textit{MPF.2021.2.8}, i.e., excluding any snapshots with a final energy per atom greater than 50 meV atom$^{-1}$ or atom distance less than 0.5 \AA, we have further fine-tuned the dataset by excluding ionic steps where: (1) electronic relaxation has not been reached and (2) at least one atom have no neighbors within the cutoff radius (5 \AA). Last but not least, data of all structures with forces over 10 eV \AA$^{-1}$~were removed or substituted with better converged PES information. (see detailed discussion in below paragraphs of DFT calculations) This cleaned-up dataset contains a total of 1,315,097 structures, and henceforth will be known as the \textit{MPF.2021.2.8.All} dataset.

\subsubsection{Ti-H dataset}
To generate a comprehensive configuration space for the Ti-H chemistry\cite{san-martinTiHydrogenTitaniumSystem1987, wipfSolubilityDiffusionHydrogen2001}, NpT MD simulations using the using the refitted M3GNet-DIRECT UP were carried out on 91 supercells of crystalline and grain boundary \ce{TiH_$n$} ($0\le n \le 2$) structures at 300, 1000 and 3000 K, and a 3$\times$3$\times$2 supercell of the most stable \ce{H2} structure in MP (P6$_{3}$/mmc, mp-24504) at 300 K. The 91 supercells of \ce{TiH_$n$} include the following:
\begin{enumerate}
    \item 36 crystalline \ce{Ti36H_{n}} supercells with H interstitial defects ($0\le n \le 72$). Four phases of Ti are considered, which are hexagonal (P6/mmm, mp-72), hexagonal close pack (hcp, P6$_3$/mmc, mp-46), body-centered cubic (bcc, Im$\Bar{3}$m, mp-73) and face-centered cubic (fcc, Fm$\Bar{3}$m, mp-6985). The latter three are known \ce{TiH_$n$} phases in experiment\cite{san-martinTiHydrogenTitaniumSystem1987}, while the P6/mmm Ti is a polymorph with comparable formation energy as that of the known HCP Ti according to the Materials Project.\cite{jainCommentaryMaterialsProject2013} For each phase, 9 compositions are spanned from $n=0$ to $n=2$, and H atoms are randomly inserted to tetrahedral and octahedral interstitial sites, following experimental observations.\cite{san-martinTiHydrogenTitaniumSystem1987}
    \item A 3$\times$3$\times$4 supercell of the face-centered tetragonal (fct) $\varepsilon$-\ce{TiH2} (I4/mmm, mp-24726), which has the maximum known hydrogen storage in Ti ($n=2$) at any temperature or pressure.\cite{san-martinTiHydrogenTitaniumSystem1987}
    \item 54 grain boundary (GB) Ti structures with $n$ H interstitial defects ($0 \le n \le 2$). For each of the three known competing \ce{TiH_$n$} phases in experiment\cite{san-martinTiHydrogenTitaniumSystem1987}, two GB orientations were considered, including one twist GB and one symmetric-tilt GB. Following the same procedure of inserting H atoms into crystalline Ti, 9 stoichiometries of \ce{TiH_$n$} ($0 \le n \le 2$) were generated for each GB orientation. In every GB model, the distance between the two boundaries is at least 10 \AA. 
\end{enumerate}
All 274 MD runs were performed to 10 ps with a time step of 0.5 fs, in line with previous AIMD works for hydrogen diffusion.\cite{tangHydrogenDiffusionPlutonium2021,tseAnalysisHydratedProton2015} Therefore, each MD trajectory contains 20,001 snapshots, where DIRECT sampling was applied to select one snapshot from each of one thousand clusters, constructing a configuration space of 274,000 MD snapshots. 

\subsection{Structure encoders}

The MatErials Graph Network (MEGNet) and Materials 3-body Graph Network (M3GNet) formation energy models trained on the 2019.4.1 Materials Project crystals data set were used as structure encoders. Both the MEGNet and M3GNet models have been described extensively in previous works\cite{chenGraphNetworksUniversal2019, chenLearningPropertiesOrdered2021, chenAtomSetsHierarchicalTransfer2021, chenUniversalGraphDeep2022}, and interested readers are referred to those publications for details. After performing graph convolutions, the output graph features are concatenated (96 elements of atomic, bond and state vectors for MEGNet and 128 elements of atomic and state vectors for M3GNet) and passed through multi-layer perceptrons to generate the final output property. The final concatenated vectors from these models therefore encode the relevant structure/chemistry for the prediction of the formation energy. In this work, the concatenated 96-D vector of MEGNet and the concatenated 128-D vector of M3GNet were utilized as structure features.

\subsection{MLIP fitting}

\subsubsection{M3GNet universal potential}

To refit the M3GNet UP, we have adopted the same settings as that used in the training of the original M3GNet UP\cite{chenUniversalGraphDeep2022}, including a 90:5:5 train:validation:test random split, a 1:1:0.1 weight ratio for energy (eV atom$^{-1}$), force (eV \AA$^{-1}$) and stress (GPa) in a Huber loss function with $\delta=0.01$, an Adam optimizer with initial rate of $10^{-3}$ and a cosine decay to $10^{-5}$ in 100 epochs. One significant modification from the original M3GNet UP is that the model complexity is expanded by doubling the dimension of both atom embeddings and multi-layer perceptrons from 64 to 128. The performance of the M3GNet UPs trained with the original model complexity is provided in Figure S2 for comparison. Further, the isolated atoms of all 89 elements in \textit{MPF.2021.2.8.All} were added into M3GNet training set to improve the extrapolability of the final potential. All other structures with isolated atoms were removed from the training set. Finally, for faster convergence, training was stopped if the validation metric did not improve for 40 epochs, instead of 200 epochs. 

\subsubsection{Moment tensor potential for Ti-H}

Two moment tensor potentials (MTPs)\cite{shapeevMomentTensorPotentials2016, gubaevAcceleratingHighthroughputSearches2019} were fitted with two training sets, i.e., AL set and DIRECT set, for the Ti-H system. The MTP cutoff radius $r_\text{c}$ and maximum level $lev_\text{max}$ were fixed at 5 \AA~and 20, respectively. In line with previous works, the weights of energies, forces and stresses were set at 1, 0.01 and 0, respectively.\cite{zuoPerformanceCostAssessment2020, liComplexStrengtheningMechanisms2020, qiBridgingGapSimulated2021} AL was conducted using the protocol developed by \citet{gubaevAcceleratingHighthroughputSearches2019} under exactly the same 274 MD scenarios explored by M3GNet-DIRECT UP. The cutoff extrapolation grade for breaking the simulation and selection of structures was set at 3, i.e., $\gamma_{break}=\gamma_{select}=3$. All 274 MD runs can reliably run for $>$ 10 ps after 15 AL iterations. All training, evaluations and simulations with MTP were performed using MLIP~\cite{shapeevMomentTensorPotentials2016,gubaevAcceleratingHighthroughputSearches2019}, LAMMPS~\cite{thompsonLAMMPSFlexibleSimulation2022} and the open-source Materials Machine Learning (maml) Python package~\cite{chenMamlMaterialsMachine2020}.

\subsection{DFT calculations}

DFT calculations were performed using the Vienna \textit{ab initio} simulation package (VASP~\cite{kresseEfficiencyAbinitioTotal1996,kresseEfficientIterativeSchemes1996}). The Perdew-Burke-Ernzerhof (PBE~\cite{perdewGeneralizedGradientApproximation1996}) generalized gradient approximation (GGA) functional was used for \textit{MPF.2021.2.8.All} and Ti-H systems.

\subsubsection{\textit{MPF.2021.2.8.All}}
Spin-polarized self-consistent calculations were carried out to all the 13,614 structures in \textit{MPF.2021.2.8.All} with forces over 10 eV \AA$^{-1}$~to obtain more accurate PES information than that of the loosely converged electronic relaxations by geometry optimizations in the Materials Project.\cite{jainCommentaryMaterialsProject2013}  The electronic convergence criterion (EDIFF) was set at 10$^{-5}$ eV, and the smallest allowed spacing between $k$ points (KSPACING) was set at 0.35 \AA$^{-1}$. All other settings were consistent with those used for static calculations in the Materials Project. The maximum number of electronic steps was set at 100. Over 83\% were successfully converged. 

\subsection{Hypothetical O- and S-containing compounds}
This dataset contains 506 and 291 O- and S-containing hypothetical materials, which were randomly selected from the $\sim$30 million hypothetical materials generated by \citet{chenUniversalGraphDeep2022}. One thousand hypothetical materials were initially selected for each group. DFT geometry optimizations were performed to those 2,000 structures using the settings for structure relaxations in the Materials Project. Only converged results were collected to be test sets for these two groups of hypothetical compounds. Subsequently, geometry optimizations were performed using M3GNet UPs with the same force convergence criterion of 0.1 eV \AA$^{-1}$, and the energy above hull ($E_{hull}$) was calculated relative to the DFT-calculated structures in the Materials Project. 

\subsubsection{Ti-H system}
Spin-polarized DFT calculations for \ce{TiH_$n$} were performed with an energy cutoff of 500 eV. Three AIMD NVT simulations were performed for three \ce{TiH_$n$} supercells, including HCP \ce{Ti36H2}, BCC \ce{Ti36H36} and FCC \ce{Ti36H72} at 1000 K, and one AIMD NpT simulation was conducted at 3000 K for FCC \ce{Ti36H72}. All AIMD simulations were conducted for 25,000 steps with a time step of 0.5 fs, in accordance with previous AIMD works for hydrogen diffusion.\cite{tangHydrogenDiffusionPlutonium2021,tseAnalysisHydratedProton2015} A single $\Gamma$ k point was used to sample the Brillouin zone.
Self-consistent calculations were performed with an electronic relaxation convergence threshold of $10^{-4}$ eV, while the density of the k grid in the reciprocal space was at least 100 /\AA$^{-3}$. The maximum number of electronic steps was set at 100.

\begin{acknowledgement}

The authors acknowledge support from Shell International Exploration and Production Inc. (Contract No. CW649697). The authors also acknowledge data and software infrastructure supported by the Materials Project, funded by the US Department of Energy, Office of Science, Office of Basic Energy Sciences, Materials Sciences and Engineering Division. Part of this work was carried out under the auspices of the US Department of Energy by Lawrence Livermore National Laboratory (LLNL) under contract No. DE-AC52-07NA27344. B.C.W. and T.A.P. acknowledge support from LLNL Laboratory Directed Research and Development (LDRD) Program Grant No. 20-SI-004 and 22-ERD-014. Computational work was performed using the Expanse at the San Diego Supercomputer Center (SDSC) from the Advanced Cyberinfrastructure Coordination Ecosystem: Services and Support (ACCESS) program supported by National Science Foundation grants No. 2138259, 2138286, 2138307, 2137603, and 2138296, in addition to the Triton Super Computer Center (TSCC) at the University of California, San Diego and the National Energy Research Scientific Computing Center (NERSC). 

\end{acknowledgement}

% \begin{suppinfo}
 
% % If you have supplementary information, most journals require you to briefly describe what's in the SI here. 

% \end{suppinfo}

\section{Data availability}
All data required to reproduce the DIRECT sampling results are available to download at: \verb!https://figshare.com/articles/dataset/20230723_figshare_DIRECT_zip/23734134!. It contains training, validation, and test data for M3GNet-MS and M3GNet-DIRECT UPs, training and test data for MTP-AL and MTP-DIRECT for Ti-H, as well as pre-processed M3GNet features of the 1.3 million structures in \textit{MPF.2021.2.8.all} and the 274,000 M3GNet MD structures of Ti-H.

\section{Code availability}
The DIRECT sampling is implemented as a scikit-learn pipeline in the MAterials Machine Learning (maml) public Github repository (https://github.com/materialsvirtuallab/maml). Example notebooks are provided in the repository to reproduce the sampling for \textit{MPF.2021.2.8.all} and M3GNet MD structures of Ti-H.

\section{Author contributions}
J.Q. generated the idea and conducted all the simulations. T.W.K helped with M3GNet model training and evaluations. S.P.O. supervised the whole project. B.C.W and T.A.P. supervised the Ti-H section. J.Q. and S.P.O. drafted the manuscript. All authors revised the manuscript.

\section{Competing interests}
All authors declare that they have no financial or non-financial competing interests. 

\clearpage

% \listoffigures
\bibliography{ref}
\end{document}

\begin{figure}[H]
    \centering
    \includegraphics[width=0.9\textwidth]{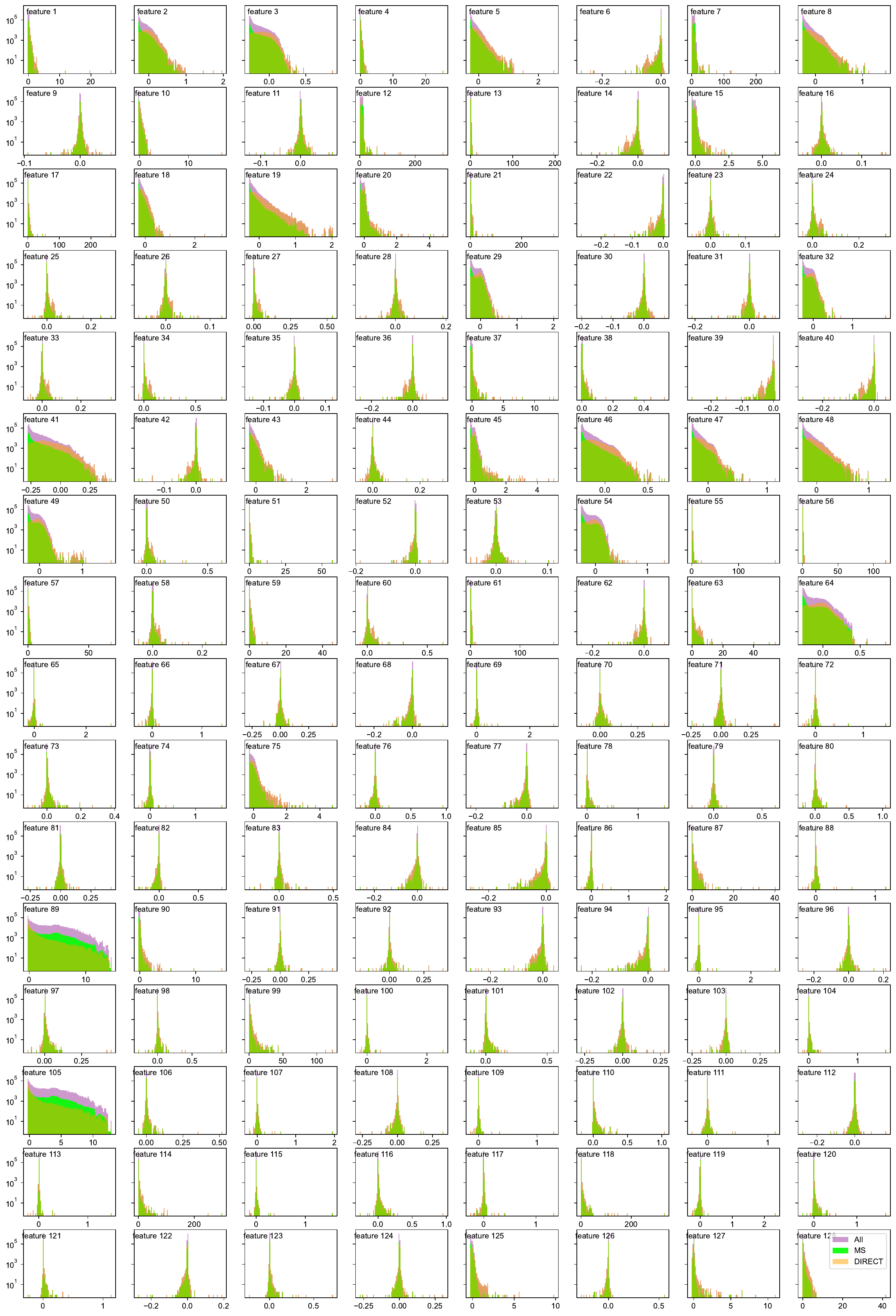}
    \caption{Distribution of all the 128 elements in M3GNet structural features of structures in DIRECT, MS and the entire MPF.2021.2.8 dataset.}
    \label{SI_fig:MP_M3GNet_feature_coverage}
\end{figure}

\begin{figure}[H]
\centering
\begin{subfigure}[b]{0.31\textwidth}
    \centering
	\includegraphics[width=\textwidth]{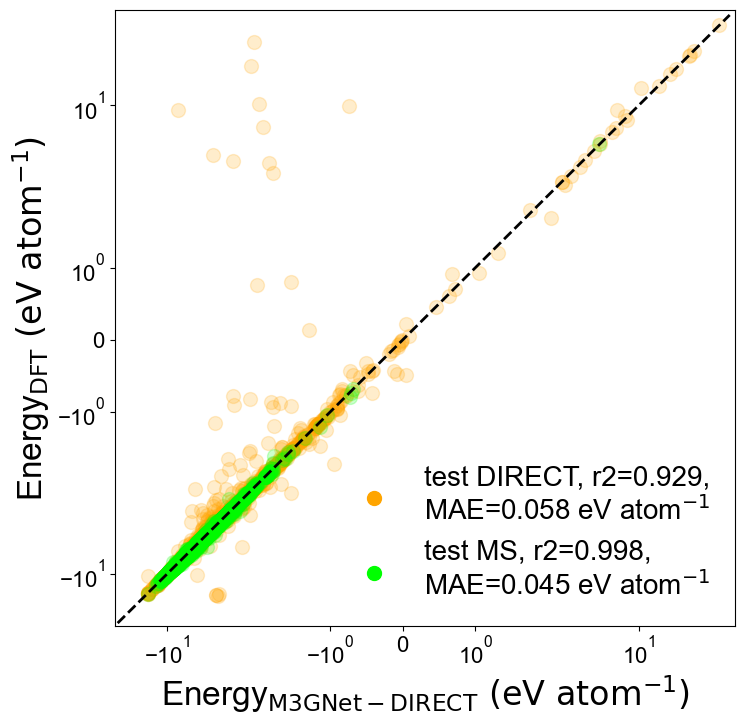}
	\caption{\label{SI_subfig:parity_M3GNet_CS_E}}
\end{subfigure}
\begin{subfigure}[b]{0.31\textwidth}
    \centering
	\includegraphics[width=\textwidth]{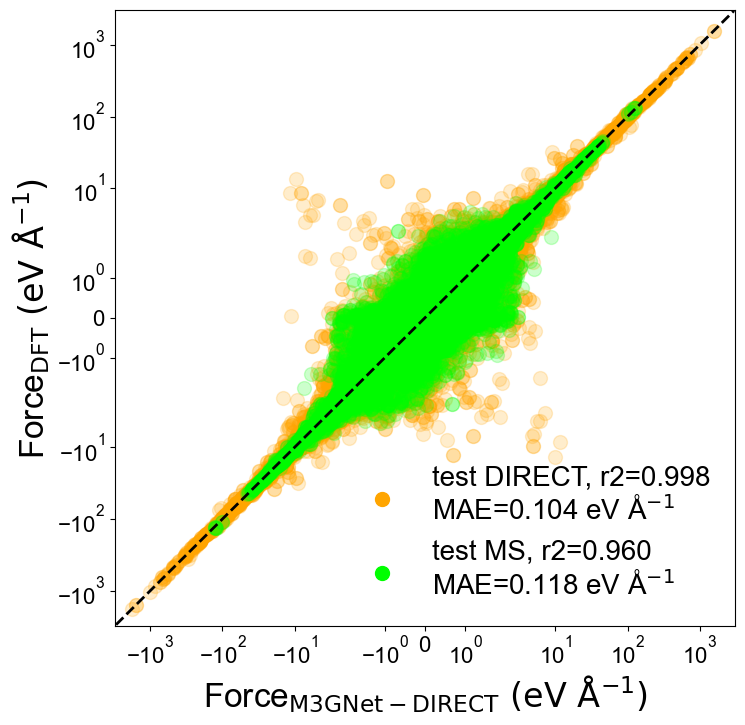}
	\caption{\label{SI_subfig:parity_M3GNet_CS_F}}
\end{subfigure}
\begin{subfigure}[b]{0.31\textwidth}
    \centering
	\includegraphics[width=\textwidth]{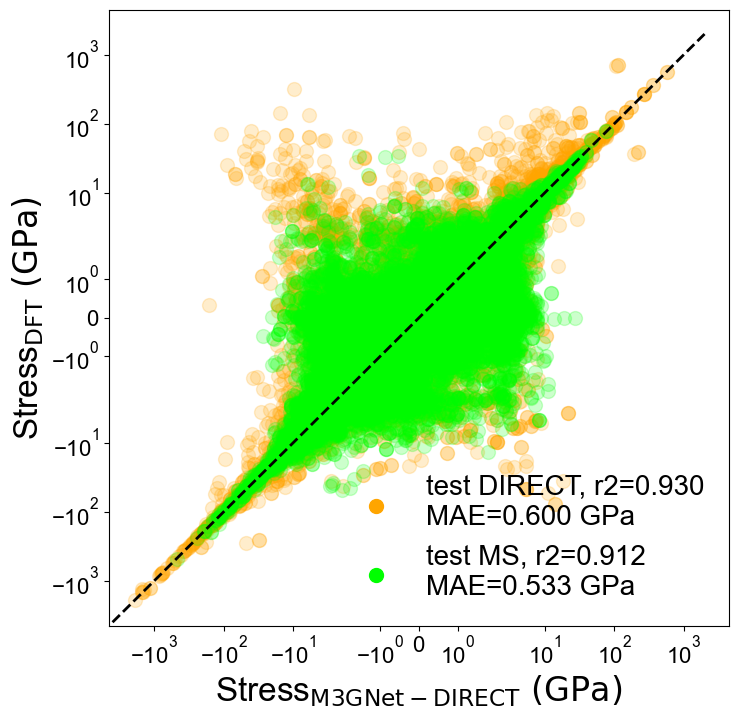}
	\caption{\label{SI_subfig:parity_M3GNet_CS_S}}
\end{subfigure}
\begin{subfigure}[b]{0.31\textwidth}
    \centering
	\includegraphics[width=\textwidth]{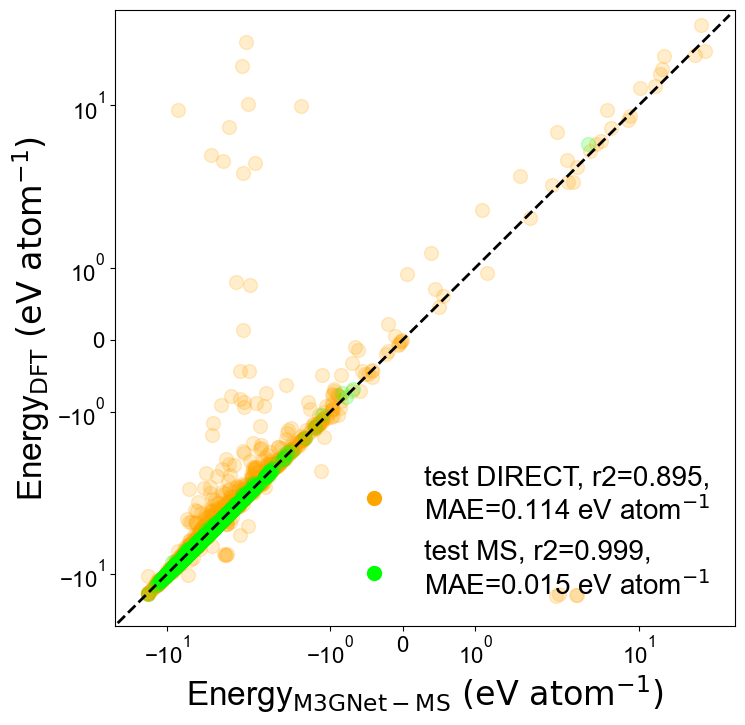}
	\caption{\label{SI_subfig:parity_M3GNet_MS_E}}
\end{subfigure}
\begin{subfigure}[b]{0.31\textwidth}
    \centering
	\includegraphics[width=\textwidth]{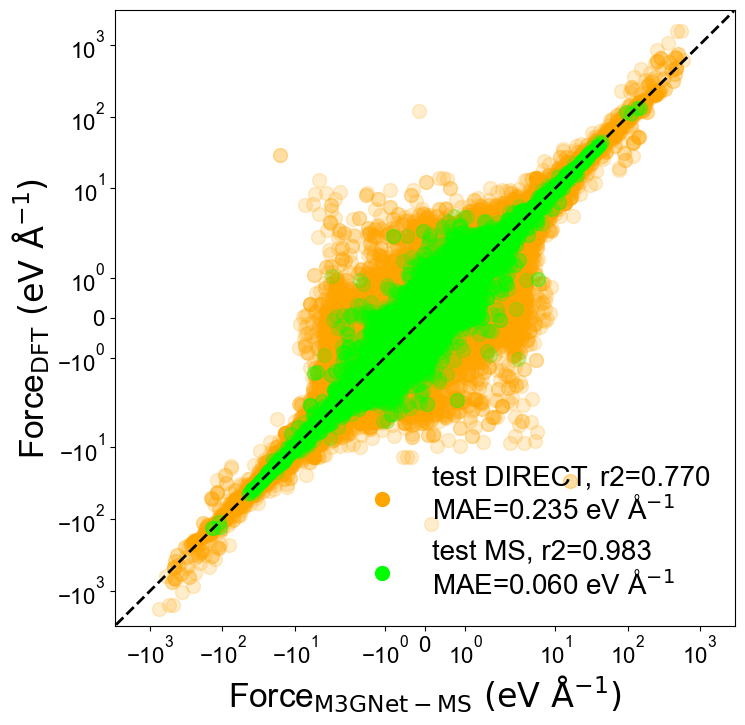}
	\caption{\label{SI_subfig:parity_M3GNet_MS_F}}
\end{subfigure}
\begin{subfigure}[b]{0.31\textwidth}
    \centering
	\includegraphics[width=\textwidth]{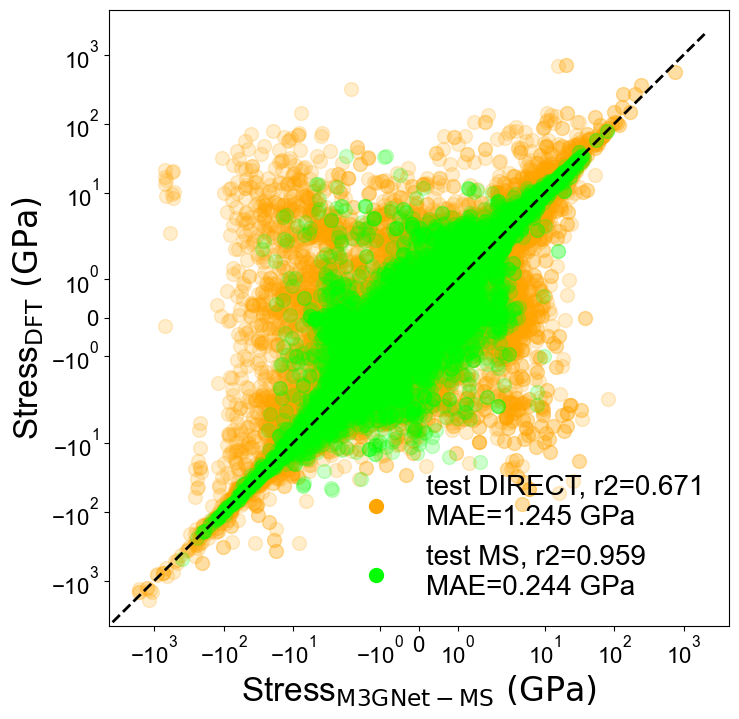}
	\caption{\label{SI_subfig:parity_M3GNet_MS_S}}
\end{subfigure}
\begin{subfigure}[b]{0.31\textwidth}
    \centering
	\includegraphics[width=\textwidth]{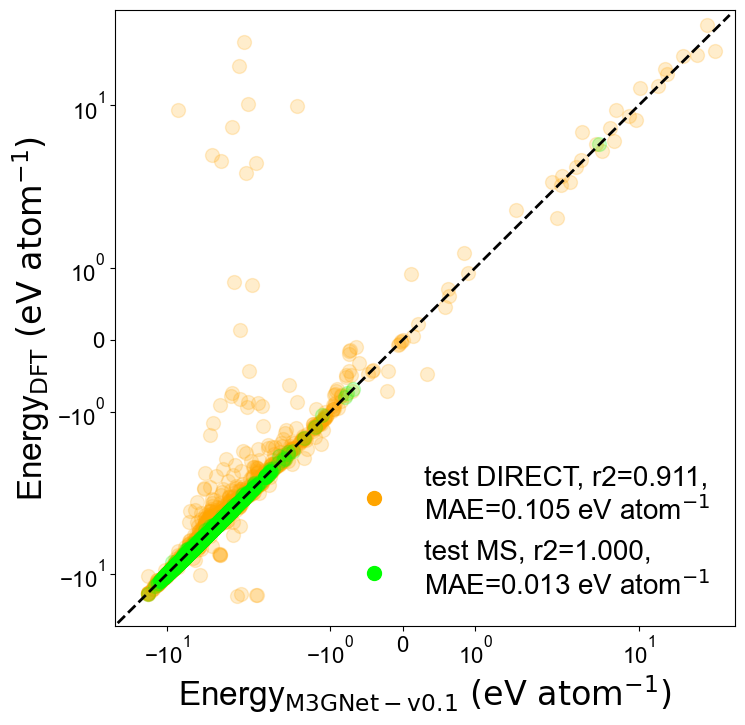}
	\caption{\label{SI_subfig:parity_M3GNet_001_E}}
\end{subfigure}
\begin{subfigure}[b]{0.31\textwidth}
    \centering
        \includegraphics[width=\textwidth]{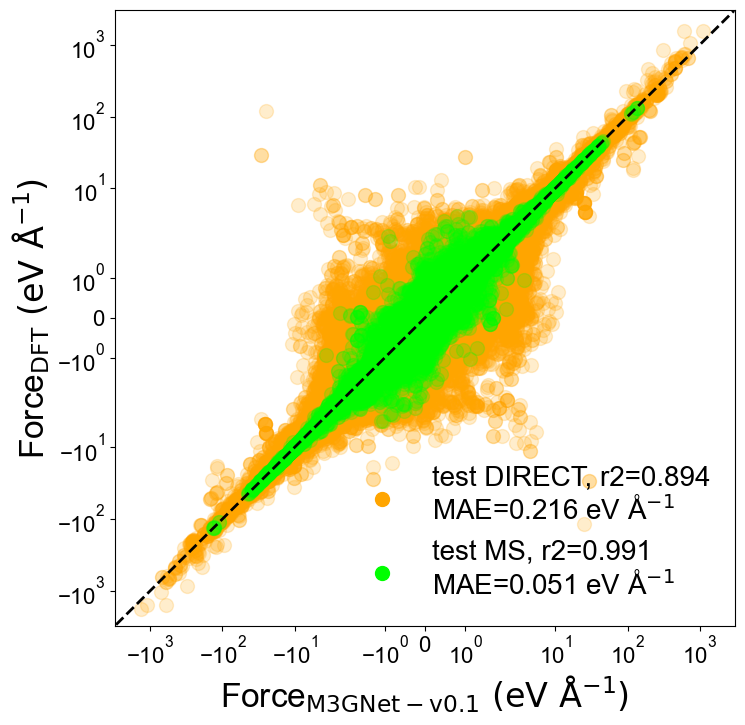}
    \caption{\label{SI_subfig:parity_M3GNet_001_F}}
\end{subfigure}
\begin{subfigure}[b]{0.31\textwidth}
    \centering
        \includegraphics[width=\textwidth]{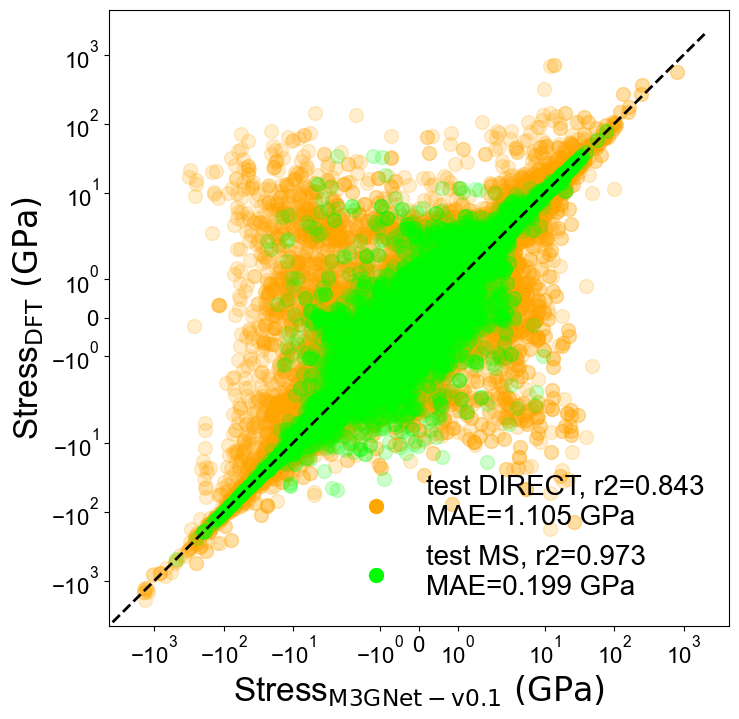}
    \caption{\label{SI_subfig:parity_M3GNet_001_S}}
\end{subfigure}

\caption{\label{SI_fig:MP_M3GNet_EFS_parity}Performance of M3GNet UPs with the same model complexity as that of the pre-trained M3GNet-v0.1 by \citet{chenUniversalGraphDeep2022}. (See details in Methods) Parity plots for (a) energies, (b) forces and (c) stresses for the M3GNet UP trained on the DIRECT set. The equivalent plots for the M3GNet UP trained on the MS set is shown in plots (d)-(f). The respective plots for the pre-trained M3GNet-v0.1 are also provided for comparison.}
\end{figure}

\bibliography{ref}